\documentstyle[aps,prb,epsfig,epsf]{revtex}
\begin{document}
\draft
\title{Fermionic SK-models with Hubbard interaction:
Magnetism and electronic structure}
\author{R. Oppermann$^{1}$ and D. Sherrington$^2$}
\address{$^1$ Institut f. Theoretische Physik, Universit\"at
W\"urzburg, Am Hubland, 97074 W\"urzburg, FRG}
\address{$^2$ Department of Physics, University of Oxford, Theoretical Physics,
1 Keble Road, Oxford OX1 3NP, UK}
\date{July 26, 2002}
\maketitle
\begin{abstract}
Models with range-free frustrated Ising spin interaction and additional Hubbard
interaction are treated exactly by means of the discrete time slicing method of
Grassmann field theory.
Critical and tricritical points, spin- and charge correlations, and the fermion
propagator, are derived as a function of temperature, chemical potential
$\mu$, of the Hubbard coupling $U$, and of the spin glass energy $J$.
$U$ is allowed to be either repulsive ($U>0$) or attractive ($U<0$).
Cuts through the multi-dimensional phase diagram are obtained.
Analytical and numerical evaluations take important replica symmetry breaking
($RSB$)-effects into account. Results for the ordered phase are given
at least in one-step approximation ($1RSB$), for $T=0$ we report the first
two-, three-, and four-step calculations ($4RSB$) for fermionic spin glasses.
The use of exact relations and invariances under $RSB$ together
with $2RSB$-calculations for all fillings and $4RSB$-solutions for
half filling allow to model exact solutions by interpolation.
For $T=0$, our numerical results provide strong evidence that
the exact spin glass pseudogap obeys $\rho(E)=c_1 |E-E_F|$ for
energies close to the Fermi level with $c_1\approx 0.13$. Rapid
convergence of $\rho'(E_F)$ under increasing order of $RSB$ is
observed and $\rho''(E)$ is evaluated to estimate subleading powers.
Over a wide range of the pseudogap and after a small transient regime
$\rho(E)$ regains a linear shape with larger slope and a small S-like perturbation.
The leading term resembles the Efros-Shklovskii Coulomb pseudogap
of two-dimensional localized disordered fermionic systems.
Beyond half filling we obtain a $\nu-1\sim(\mu-U)^2,\mu\geq U,$ dependence of the
fermion filling factor $\nu$.
We find a half filling transition between a phase for $U>\mu$, where the Fermi level
lies inside the Hubbard gap, into a phase where $\mu(>U)$ is located at the center of the
upper spin glass pseudogap (SG-gap). For $\mu>U$ the Hubbard gap combines with the
lower one of two SG-gaps (phase I), while for $\mu<U$ it joins the sole SG-gap which
exists in this half-filling regime (phase II).
Shoulders of the combined gaps are shaped by $RSB$ due to spin glass order.
We predict scaling behaviour at the half filling transition which becomes continuous
due to $\infty RSB$.
Implications of the half-filling transition between the deeper insulating phase II
and phase I for the eventual delocalization by additional hopping processes in
itinerant model extensions are discussed.
Possible metal-insulator transition scenarios are described.
Generalizations to random Hubbard coupling and alloy models as well as
frustrated magnetic interactions with ferro- or antiferromagnetic components are
also considered separately.
\end{abstract}
\section{Introduction}
The interplay between complex magnetic order and electronic properties including
transport attracts growing interest despite an already long history of active and
successful research in this field.
Reviews of many physical phenomena which result from competing interactions
in frustrated magnetic systems can be found for example in Ref.\onlinecite{Diep.book}
while coupling with electronic transport is discussed in Ref.\onlinecite{fisherhertz}.
The particular role of Hubbard interactions in the context of glassy order was
considered for random transition metal alloys in Ref.\onlinecite{mihill.shrgtn}
already more than three decades ago.
This work described the multiple competition between different interactions,
which link magnetism with electronic transport such as mobile carriers in
contact with spin glass order, and noted also the interference of the
Kondo effect. Moreover an analogy with Anderson localization theory was formulated.
At that time the theory of Anderson localization continued to progress over
many years and was finally shaped in its replicated field theoretic version in the
early eighties, while, independently and almost simultaneously with this event,
Parisi discovered the highly complex replica symmetry broken solution of
range-free spin glass order \cite{parisi,parisibook}.
These theories remained remarkably disconnected for a long time, if one disregards
for example the analogy discussed in Ref.\onlinecite{mihill.shrgtn}.
On one hand, one knew that the unitary universality class of Anderson localization
takes into account broken time-reversal invariance in the case of random magnetic
scattering centers. However this class, derived for noninteracting disordered systems,
cannot simply be assumed to describe localization of charge carriers due to their
exposure to spin glass order, which itself is an effect of magnetic interaction.
While the potential power of random magnetic order to localize charge carriers is
evident, the specific part played by the many body interactions, which can create gaps
in the density of states near the Fermi level and other complicated fluctuation
effects, must also be taken care of. A well-known case where disorder and interaction
are essential is that of Mott-Anderson transitions\cite{finkelstein}. The
role of magnetic fluctuations due to singularities in the triplet
channel remain a matter of concern.
In recent years, the Mott transition of the Hubbard model, ignoring the
long-range Coulomb interaction, raised a lot of interest too
\cite{kotliar.et.al,balachandran.review}.
The theory of insulators and in particular of antiferromagnetic materials is to a
large extent connected with the Hubbard model. The apparent link between magnetism
and transport properties and the presence of a metal-insulator transition renders the
model particularly attractive. Far from simple however is the many body theory and
even the mean field theory of the Hubbard model\cite{vollhardt}. Expansions around
its local, so-called atomic, limit have been elaborated by Metzner \cite{metzner}.
The necessity to generate the effective magnetic interactions and to calculate their
magnetic effects in cooperation with transport properties on the basis of the simple
atomic limit requires a lot from a perturbation technique.
Effective magnetic interaction models were also derived, for example the famous
$t-J$ model, which project onto a subset of relevant variables.

An example which combines, on the same footing, the Hubbard interaction with magnetic
effects of a (nonlocal) spin interaction, was introduced in Ref.\onlinecite{laloux}
and found necessary in order to deal with weak localization in $He^3$. \\
Our present paper introduces and analyzes a model defined by the grand-canonical
Hamiltonian
\begin{equation}
{\cal{H}}=U\sum_i n_{i\uparrow}n_{i\downarrow}-\sum_{(ij)}J_{ij}\sigma_i
\sigma_j -\mu\sum_{i\sigma}n_{i\sigma}
\label{JU}
\end{equation}
where the $(ij)$ run over all pairs and the random magnetic interaction involves
Gaussian distributed independent couplings $J_{ij}$. Spin and
charge operators $\sigma_i=n_{i\uparrow}-n_{i\downarrow}$ and $n\equiv
c^{\dagger}c$ are given in terms of fermion operators $c$. The
Hubbard interaction is nonrandom (except in a final section
\ref{sec:randomU}).
The model aims at the description of insulating phases, where the leading role is
played by the cooperation of Hubbard- and {\it frustrated} spin-interactions.
One can view the model as the {\it localized limit of a Hubbard model with frustrated
spin interactions} and hence, in addition, as a starting point with highly nontrivial
magnetic behaviour for expansions with respect to electron hopping.
We show that the model is soluble for range-free $J_{ij}$ to the extent that one can
derive the full replica symmetry broken solution. \\
The study of such insulating models, or of the insulating side of itinerant models,
helps to analyze the precursor of the onset of transport-related phenomena which
depend on the interplay between magnetism and energy gaps (in particular at the
Fermi level). The breakdown of magnetism and a subsequent and related filling
of the gap, when hopping is added, can induce an insulator metal transition.

Another goal of the research presented in this paper is to study the enhancement
of insulating properties and the transition between distinguishable types
of insulating phases. The enhancement we find here originates in a constructive
superposition of gap widths.
It is in general of considerable technological relevance to identify all elementary
interaction mechanisms which reinforce or change in a controlled way the insulating
tendency of materials and, at the same time, to find relationships with magnetic order.
Random magnetic interactions deepen insulating properties by strong density of states
depletion near the Fermi level and in addition by the randomness which generates
a different type of Anderson localization.

The present work reports progress in understanding insulating fermionic models
which incorporate Hubbard interactions in the fermionic extension of the
range-free Sherrington-Kirkpatrick spin glass interaction ($SK$-model)
\cite{fisherhertz,SK,binderyoung}.
We note in passing that the model contains as special cases not
only the standard $SK$-model (at imaginary chemical potential
$\mu=i \pi T/2$) and the atomic limit of the Hubbard model, but
also the Ghatak-Sherrington (spin 1) model \cite{ghatak} and of
course, for vanishing Hubbard interaction, the fermionic
$SK$-model, which was analyzed in Ref.\onlinecite{rohf} on
1$RSB$-level. When a strong Hubbard repulsion effectively
evacuates double occupied sites the model operates in an almost
three-state per site space. In this regime it is also similar to
the model of randomly placed strong $U$-scattering centers
\cite{hertz} provided the fermions are localized.

The article is organized in three main parts $A-C$ and
two additional ones $D,E$. Part $A$ is purely analytical, while
$B$ and $C$ contain the numerical analyses of the exact selfconsistent
equations obtained from $A$. In particular, the reader will find in

${\bf A}$) [section \ref{sec:technique}] the derivation of the analytic solution by
means of the discrete time slicing method, where all fields representing the Hubbard
interaction are integrated out in perfect coexistence with the spin glass related
fields,

${\bf B}$) [sections \ref{section:Tf}-\ref{section:2RSB}]-
a detailed numerical analysis of critical and tricritical temperatures,
cuts through the multi-dimensional phase diagram,
solutions of the coupled selfconsistent integral equations in the ordered
phase, their subsequent use in the calculation of the fermion propagator
including the density of states and a calculation of the fermion
concentration, all given as a function of the chemical potential
$\mu$.
The full range of $\mu$ for which the spin glass phase can exist
is covered.
Particular attention is paid to the large replica symmetry breaking ($RSB$-)
effects at low temperatures including $T=0$. For $T=0$ we present the first
2-step $RSB$-evaluation for fermionic spin glasses (with arbitrary filling),

${\bf C}$) [section \ref{section:4RSB}] the 4-step $RSB$ solutions
for half filling which lead, by interpolation with known exact relations,
to a prediction of the exact form of the spin glass pseudogap.
The nonanalytic behaviour is determined and a comparison with the
Efros-Shklovskii Coulomb pseudogap included.

In the remaining smaller part $D$ randomness in the Hubbard coupling is
included in order to control changes in the domain of discontinuous and continuous
phase transitions, and finally in $E$, implications of the gap structure derived
in part $B$ are discussed. In particular we include in

${\bf D}$) [section IX] a phenomenological discussion of the onset of
diffusive transport beyond a metal-insulator transition. Two scenarios for
metal-insulator transitions governed by the random and frustrated magnetic
interaction of spin glasses are described. This part serves to describe one of the
major aims of our current research. Finally in

${\bf E}$) [section X, XI] the effect of several distributions of random Hubbard
interaction on the freezing temperature $T_f$ are analyzed, and a discussion of
ferro- or antiferromagnetic effects induced by nonzero mean values of $J_{ij}$
concludes the paper.

Continued efforts to approach the (best possible) analytical $T=0$-solution for the
mean field theory of non-itinerant spin glasses form a part of the paper.
Sufficiently good qualitative, if not quantitative, knowledge of this localized
$T=0$ solution seems necessary, or at least a helpful basis for the analysis of
itinerant glassy magnetism in fermionic systems. This can easily be understood
if one reviews the strong low temperature $RSB$-effects
on relevant properties (like shape, effective width, and stability against
thermal filling) of gaps in the band structure. Pseudogaps at the Fermi level or
smallness of the density of states in the vicinity of $E_F$ are features which
undoubtedly are relevant for two-particle correlation functions such as
conductivities.
For this reason $RSB$ has the power to influence $T=0$ phase transitions and
must be controlled, in particular with respect to preditions for
metal-insulator transitions. It is also interesting to compare our results with
spin dynamical fluctuation effects in disordered or translationally invariant
quantum-spin systems.
Unfortunately (in the sense of solvability) $T=0$ phase transitions often occur
as quantum spin glass transitions which involve utterly complicated connections
between quantum-spin dynamics, critical behaviour and $RSB$.

Implications of $RSB$ in many body quantum physics have not yet been sufficiently
explored. We mention two examples:
i) localization theory involves non self-averaging quantities and a $RSB$-free theory
is questionable with respect to these observables, and
ii) recent theories of quantum critical points in low-dimensional transverse field
Ising systems have shown the importance of infinite-disorder fixed points
\cite{huse}, which were expected to be applicable to quantum spin glass transitions
as well. Their relevance in high dimensions (domain of attraction) and even in
mean-field limits must be analyzed.
Although the comparison between non-replicated and replicated methods is often
difficult, the claim of relevance of strong coupling theories can well be related
to the strong coupling theory needed to incorporate complete $RSB$, which was a
motivation of the present research.
We cannot answer these questions in the present paper, but our attempt to find
analytical solutions at $T=0$ for the present infinite-range problem is also meant
as a step towards finite-dimensional systems and the control of $RSB$-relevance.
If one wants to understand the term 'relevance' in the renormalization-group sense,
it is important that continuous $T=0$ phase transitions can exist in the system.
While the magnetic breakdown is discontinuous in the mean field case and probably
in general too, we derive and emphasize the special nature of half-filling
transitions of the present model. Moreover this transition is sharp only at $T=0$
and becomes continuous only due to infinite $RSB$, while in other cases $RSB$ renders
a first order transition even more discontinuous (one example given in
Ref.\onlinecite{osk}).

Also recently and during progress of the present work, a numerical approach using
infinite $RSB$ equations \cite{crisanti} was proposed which employs an interesting
analogy between the 3SAT-optimization problem \cite{crisanti,remi} and the
standard $SK$-model at $T=0$.
In our approach, apart from considering more general model, we follow a different
approach which attempts to evaluate the maximum feasible number of integrations
involved in $\infty RSB$ equations. In this paper the highest accuracy is so far the
$2RSB$-approximation of the exact solution for arbitrary values of the chemical
potential and $4RSB$ for particle hole symmetry $\mu=U/2$ (half filling).

Since the techniques applied in section \ref{sec:technique} are readily extended to
include statistical distributions of the Hubbard interaction and/or different
interaction strength for $A$ and $B$-type atoms in A-B alloys we extended our
analysis to study models described by the Hamiltonian ${\cal{H}}+\delta{\cal{H}}$
with
\begin{equation}
{\cal{\delta H}}=\sum_i \delta U_i n_{i\uparrow}n_{i\downarrow},
\label{H:randomU}
\end{equation}
where $\delta U$ obeys a wide variety of different distributions.
We evaluate changes in critical temperatures and their effect on the phase diagram.
This part helps to gain insight into the competition between repulsive and attractive
interactions on one side and spin glass order on the other.
The strong asymmetry between positive and negative $U$ are incorporated in one
model and weighted by $U$-distributions. A distribution-invariant point is found
(for symmetric distributions), i.e. a point where the critical temperature does not
depend on $U$-fluctuations at all.

Due to the absence of spin- and charge-quantum dynamics (in the hopping free model)
one may expect the model to be more easily solvable than the Hubbard model, the
solution of which faces the high obstacle of complicated dynamics. However randomness
of the magnetic interaction introduces the complication of replica symmetry breaking
at all temperatures below freezing and in particular at $T=0$ too.
The additional introduction of hopping leads to further complications,
since all problems of the Hubbard model are then contained too.
Let us mention the two most important classes of itinerant model extensions.
These are given by the single-fermion and (fermion)-pair hopping Hamiltonians by
\begin{equation}
a) \hspace{.4cm}{\cal{H}}_t=\sum_{(i,j)}t_{ij}
c^{\dagger}_{i\sigma}c_{i\sigma}\hspace{.2cm};
\hspace{.6cm}
b) \hspace{.4cm}{\cal{H}}_t^{pair}=\sum_{(i,j)}t^{(p)}_{ij}
c^{\dagger}_{i\downarrow}c^{\dagger}_{i\uparrow}c_{j\uparrow}c_{j\downarrow},
\label{hopping}
\end{equation}
which respectively have the tendency to delocalize single fermions or pairs of
fermions. Each of these additional hopping Hamiltonians can be combined with
models (1) and (2).
In section \ref{sec:MIT} we describe qualitatively effects of the single-fermion
hopping term ${\cal{H}}_t$. A perturbation theory in the fermion hopping elements
requires detailed (if not exact) knowledge of the fermion propagator in the
insulating limit, which is analyzed in section VIIIC.

The pair-hopping Hamiltonian ${\cal{H}}^{pair}_t$, on the other
hand, can cause a coherent pair state with various consequences.
Related questions concern the possible existence of microscopic superconducting
glasses. An interesting simplification of 3b) if compared to 3a) is the fact that
$H_t^{pair}$ commutes with Hamiltonian (\ref{JU}), which means that the spin
glass field $Q$ has a static saddle point. Quantum dynamics of the $SU(2)$ generators
$c_{\uparrow}c_{\downarrow}$, $c^{\dagger}_{\downarrow}c^{\dagger}_{\uparrow}$ does
not invalidate a $Q$-static theory (unlike the $H_t$ Hamiltonian). Since the pair
operators are $SU(2)$ generators as well, a distribution of the pair-hopping elements
analogous to the $J_{ij}$ in an SK-model would generate an $XY$ superconducting glass
model. Its description requires a quantum-dynamic order parameter and thus falls into
another class of low temperature behaviour.

Experiments on two-dimensional electron gases ($2DEG$) with strong ferromagnetic
coupling to local moments \cite{awschalom,awschalom2} demonstrated the relation
between transport and magnetism in the presence of disorder and correlation.
A spectacular giant magnetoresistance was observed in small fields while a transition
into a quantum Hall liquid took place at high fields. The 2D-geometries were
intentionally used to rule out orbital effects and in Ref.\onlinecite{awschalom}
to suppress magnetic order. Results of our paper show that a strong ferromagnetic
coupling to a spin glass ordered II-VI semiconductor, for example, would present a
novel case with certain new fundamental questions reaching beyond the mere
comparison of the difference between paramagnetic- and spin glass response. This
experimental situation would test localization mechanisms of more than one kind (and
different from the unconfirmed \cite{awschalom} possibility of polaron formation):
given that the strong coupling between $2DEG$ and randomly oriented magnetic moments
support Anderson localization while a spin glass pseudo-gap, which we discuss
in this paper in comparison with the Efros-Shklovskii gap, is expected to contribute
a particular stretched-exponential low temperature resistance. Moreover the spin
glass order does not smear the correlation-induced level correlation, but increases
it further as described below.

One final remark refers to the energy scale used in our article: since we focus on the
spin glass phase generated by a Gaussian-distributed random spin interaction the
most important energy scale is set by
$J=\sqrt{\langle(J_{ij}-\langle J_{ij}\rangle)^2\rangle}$.
{\it We work hence in $J$-normalized dimensionless energy variables} in order to
ease reading. Only when needed, this normalization is mentioned explicitly.
\section{Exact derivation of selfconsistent schemes for frustrated range-free
magnetic interaction and Hubbard interaction}
\label{sec:technique}
Working the double-interaction model (\ref{JU}) into a replicated Grassmann
field theory is a standard procedure. The Grassmann field technique is explained
for example in Ref.\onlinecite{negeleorland}.
Usually a continuous time limit is taken before the fermionic fields are integrated
out. This involves an unknown infinite constant which is removed by a regularization.
The discrete time--slicing method can instead be used and all integrations performed
before the continuum time limit is taken at the end.
This is known to be the exact procedure of field theory.
It avoids singularities of the symbolic continuous--time formalism
and hence does not require regularization. Its feasibility depends on whether a
quadratic form in the Grassmann fields can be obtained and in particular be
diagonalized. In spite of an $O(\psi^8)$ effective spin glass interaction
(before decoupling) and the subsequent complication by $RSB$, we found it convenient
to solve the problem in this exact way (up to the analysis of $RSB$ which is the
only source of approximations in the paper).

The Hubbard interaction, local in real space and in time, competes
with the infinite range SK Hamiltonian and its statistical spin correlations,
which are local in time but also involve time--independent (infinite-ranged in time)
magnetic disorder correlations. Both interactions can be decoupled altogether.
Not only the Grassmann integrals can be performed but also the Hubbard-decoupling
fields will be integrated out exactly. This procedure is applied first to the
partition function in \ref{subsec:Z} and subsequently to the fermion propagator in
\ref{subsec:genfunct}, where it requires a modification.
This amounts to a diagonalization of the problem, exempting only
the final solution of infinite-step replica symmetry breaking.
Reaching a four-step accuracy at $T=0$ in section \ref{section:4RSB}
turns out to be sufficient for drawing the physically relevant conclusions.
%
\subsection{The disorder averaged grand canonical partition function and
thermodynamic behaviour}
\label{subsec:Z}
The grand canonical partition function of model (\ref{JU}) for fermions in a
particle bath with chemical potential $\mu$ can be expressed in terms of the
anticommuting eigenvalues $\psi^a$, $\bar{\psi}^a$ of the fermion operators
$c$, $c^{\dagger}$. This is achieved by inserting the representation of $1$ in
terms of fermionic coherent states at $M-1$ equidistant time slices of the imaginary
time interval $0\leq\tau\leq\beta\equiv1/(k_B T)$.
The index $a$ denotes an n-fold replication of the system.
Then, as usual in the framework of the replica trick, the averaged free energy
$F\equiv -T \langle lnZ\rangle =lim_{n\rightarrow0}\langle(1-Z^n)/n\rangle$
is calculated from the disorder average of
\begin{eqnarray}
Z^n&=&\lim_{M\rightarrow\infty}\prod_{k=0}^{M-1}\prod_{i,a,\sigma}
\int
d\bar{\psi}_{i\sigma}^{a}(\tau_k)d\psi_{i\sigma}^a(\tau_k)\hspace{.1cm}
exp\left[-\epsilon\sum_a\sum_{k=0}^{M-2}
\left({\cal{K}}^a({\bar{\psi}^a_{i\alpha}(\tau_{k+1}),\psi^a_{i\alpha}(\tau_k)})
+{\cal{K}}^a(-\bar{\psi}^a_{\alpha}(\tau_0),\psi^a_{\alpha}(\tau_{M-1}))\right)\right]
\nonumber\\
&&exp\left[-\sum_{i,a,\alpha}\sum_{k=0}^{M-2}\bar{\psi}^a_{i\alpha}(\tau_{k+1})
(\psi^a_{i\alpha}(\tau_{k+1})-\psi^a_{i\alpha}(\tau_k)) -
\bar{\psi}^a_{i\alpha}(\tau_0)(\psi^a_{i\alpha}(\tau_0)
-\psi^a_{i\alpha}(\tau_{M-1}))\right]
\label{partition.function}
\end{eqnarray}
where
\begin{eqnarray}
{\cal{K}}^a(\bar{\psi},\psi)&=&\sum_{(ij)}J_{ij}\sum_{\sigma,\sigma'}
\bar{\psi}^a_{i\sigma}(\tau_{k+1})\sigma\psi^a_{i\sigma}(\tau_k)
\bar{\psi}^a_{j\sigma\prime}(\tau_{k+1})\sigma'
\psi^a_{j\sigma'}(\tau_k)+\nonumber\\
& &\sum_{i}\frac{U_i}{4}\left[(\sum_{\sigma}\bar{\psi}^a_{i\sigma}
(\tau_{k+1})\psi^a_{i\sigma}(\tau_k))^2-(\sum_{\sigma}
\bar{\psi}^a_{i\sigma}(\tau_{k+1})\sigma\psi^a_{i\sigma}(\tau_k))^2\right]-
\mu\sum_{i,\sigma}\bar{\psi}^a_{i\sigma}(\tau_{k+1})\psi^a_{i\sigma}(\tau_k).
\end{eqnarray}
The disorder average, performed over the chosen Gaussian distribution of
frustrated magnetic couplings $J_{ij}$ (but fixed Hubbard-coupling $U$ of
arbitrary sign) leads
to an effective functional with
8-fermion correlation. This integration and the subsequent decoupling of
the 8-fermion interaction (which is equivalent to the usual 4-spin interaction)
are standard and not repeated here in detail.
We just mention that the Gaussian decoupling by means of matrix fields
$Q^{ab}_i(\tau_k,\tau_{k'})$
initially depend on two times as well as replica labels, since
those carry along all degrees of freedom as obvious from the coupling term
$Q^{ab}(\tau_k,\tau_{k'})
\sum_{\sigma}\bar{\psi}^a_{\sigma}(\tau_k)\sigma\psi^a_{\sigma}(\tau_k)
\sum_{\sigma'}\bar{\psi}^b_{\sigma'}(\tau_{k'})\sigma'\psi^b_{\sigma'}(\tau_{k'})$.
The decoupling fields for the Hubbard interaction, expressed above in terms of
charge and spin operators, are dynamic; we denote them by $\alpha_k$ and $\gamma_k$
for each time instant $\tau_k$.
The spin glass $Q$-fields can in general also show quantum dynamical behaviour.
This would be the case if transport processes were allowed.
As in quantum Heisenberg spin glasses \cite{subir} or e.g. in the
famous Hubbard model in infinite dimensions \cite{kotliar.et.al}, quantum dynamics
poses even a major obstacle in finding an exact analytical mean-field solution.
We remark that the present techniques in principle allow to include quantum spin-
or charge- dynamical effects but exact solutions may not be found even in the
mean field limit.
Here we consider only cases in which the mean of $J_{ij}$ is zero
or positive so that in the following decomposition
\begin{equation}
Q_i^{ab}(\tau_k,\tau_{k'})
=q^{aa}(\tau_k-\tau_{k'})\delta_{ab} +
q^{ab}(1-\delta_{ab})+\delta Q^{ab}_i(\tau_k,\tau_{k'}).
\end{equation}
a saddle point matrix without spatial variation applies, and the site $i$ index
has been dropped in its elements $q$.
The spatial dependence of the saddle point solutions must instead be retained in
cases like glassy antiferro- or ferrimagnets, two-component magnets \cite{osk} or
similar phases, where regular spatial structures are not removed
by averaging over disorder.
In the absence of quantum spin dynamics, as is the case for the present
non-itinerant model, the spin glass order parameter fields
$Q_{k k'}\equiv Q^{ab}_i(\tau_k,\tau_{k'})$ have a purely static saddle
point. Thus one can take advantage of the time independence of
$q^{aa}(\tau_k-\tau_{k'})=\tilde{q}$, noting in addition that
$q^{ab}=\langle\langle\sigma^a\rangle\langle\sigma^b\rangle\rangle_{_{dis}}$
is always static unless non-equilibrium dynamics is taken into account.
Moreover contributions from $\delta Q$ vanish in the thermodynamic
limit. It is worth noting that also in the finite-range extension of the
present model a non-dynamic $\langle Q^{aa}\rangle$ is obtained.
In contrast to Heisenberg spin glasses or metallic spin glasses,
where the so-called static approximation of $Q^{aa}(t,t')$ is in principle inadequate
and perhaps fails to detect important features, the present model imposes a static
average of the diagonal Q-fields - we emphasize that, within the present paper,
approximations are only applied in the context of a finite number of
replica symmetry breaking steps. At present we implicitly keep
complete $RSB$ by means of the Parisi matrix $Q_{Parisi}$ and
continue to perform exact operations on the replicated partition function,
which is obtained after averaging (\ref{partition.function}) over
a Gaussian distribution of the random magnetic interaction $J_{ij}$ as
\begin{equation}
<Z^{(n)}> = const\hspace{.2cm} e^{-\frac{1}{4}N(\beta J)^2
Trace{Q_{parisi}^2}}
\left[\left[\prod_{\{r,b_r\}}\int_{{z^{^{(b_r)}}_{r}}}^G\right]
\left[\prod_{a,k}\int_{y^a}^G\int_{\alpha_k^a}^G\int_{\gamma_k^a}^G\right]
\int{\cal{D}}\Psi\hspace{.2cm}
e^{-\hspace{.0cm}{(\overline{\underline{\Psi}}}
\hspace{.1cm}{\underline{\underline{B}}}\hspace{.1cm}
{\underline{\Psi}})}\right]^N.
\label{avZ2}
\end{equation}
The superscript $G$ stands for normalized Gaussian integrals and $\{r,b_r\}$
runs over all spin decoupling fields $z_r^{^{(b_r)}}$ on which the effective field
$H_{eff}$ depends.
The spin-fields $z_r^{^{(b_r)}}$ serve to decouple all $O(\psi^4)$-terms of the form
$(\bar{\psi}\sigma\psi) Q_{Parisi}(\bar{\psi}\sigma\psi)$, which are associated with
the r-th level block matrices aligned along the diagonal of the Parisi $Q$-matrix
\cite{parisi,parisibook}.
The crucial tree structure of Parisi-$RSB$ is thus kept track of in the exponent
\begin{equation}
({\bar{\Psi}}B\Psi)\equiv\prod_{r=0}^{\infty}\sum_{b_r=b_{r+1}m_{r+1}/m_r+
(1-m_{r+1}/m_r)}^{b_{r+1}m_{r+1}/m_r}\bar{\psi}^{a}_{\sigma,k}
\left[B^{aa}_{\sigma\sigma}(H_{eff}(y^a,\{z_{\kappa}^{(b_{\kappa})}\})
\right]_{kk'}\psi_{\sigma,k'}^a\hspace{.6cm},a\equiv b_0
\end{equation}
where $\prod\sum$ symbolizes a multiple sum over all
$b_0,b_1,b_2,...,b_{\infty}$ with
boundary conditions $m_0=1$ and $m_{\infty}=n$. Integration over Grassmann fields
$\psi$ will readily yield the determinant of $\underline{\underline{B}}$.
This matrix is given by
\begin{eqnarray}
(B^{aa}_{\sigma\sigma})_{_{k+1,k}} &=&
-(1+\epsilon\hspace{.1cm}\mu+i\sqrt{\epsilon\hspace{.1cm}U/2}\hspace{.1cm}
\alpha^a_k+ \sigma\left[\sqrt{\epsilon\hspace{.1cm}U/2}\hspace{.1cm}
\gamma_k^a+H_{_{eff}}(y^a,
\{z_{_r}\})
\right])\\
(B^{aa}_{\sigma\sigma})_{_{k,k}} &=& 1\quad\quad {\rm and\hspace{.1cm}
boundary\hspace{.1cm}terms}\hspace{.3cm} B_{0,M-1}=B_{M,M-1}\hspace{.5cm},
\label{BMatrix}
\end{eqnarray}
where
\begin{equation}
H_{_{eff}}(y^a,
\{z_{_r}\})
=\sqrt{\tilde{q}-q_1}\hspace{.2cm}y^a+
\sum_{r=1}^{\infty}\sqrt{q_r-q_{r+1}}\hspace{.2cm}z_r^{(b_r)}
\label{def:H}
\end{equation}
represents the spin glass effective field. The effective field
acquires an additional magnetization contribution
$J_0 <\sigma>$ in case one allows a finite mean value of $J_0\equiv<J_{ij}>$.
Truncating the sum over $r$ after the $\kappa$-th term with $q_{\kappa+1}=0$
represents the $\kappa$-step approximation to full $RSB$, which we call $\kappa RSB$
and $\infty RSB$ respectively. The Gaussian integral operator representation
\begin{equation}
\int_z^G\equiv\int_{-\infty}^{\infty}\frac{dz}{\sqrt{2\pi}} e^{-z^2/2}
\end{equation}
provides more transparency especially in reading multiple
integrals. We use this definition everywhere in the paper.\\
For each replica index $a$, the Grassmann integration leads to the determinant
\begin{equation}
det B^{aa}_{\sigma\sigma}=1+\prod^{M-1}_{k=0}\left[1+\epsilon\hspace{.1cm}\mu
+\sqrt{\epsilon\hspace{.1cm}U/2}\hspace{.1cm}(i \alpha^a_k+\sigma\gamma^a_k)+
\epsilon\hspace{.1cm}\sigma H_{eff}(y^a,\{z_r\})\right]
\end{equation}
which still depends on the z(-spin)-field configuration.
Time slicing of the imaginary time interval $0<\tau<\beta$
was performed in $M-1$ equidistant steps of size $\epsilon=\beta/M$.
The continuum time limit and hence the exact solution will result
by taking the limit $M\rightarrow\infty$ in the end.
Let us further abbreviate by use of the shorthand notation
\begin{eqnarray}
& &\int^G_{\alpha,\gamma} det B\quad\equiv\quad
\prod_a\prod_k\int^G_{\alpha^a_k}\int^G_{\gamma^a_k}
det B\nonumber\\
&=&
(1+\prod_k\left[1+\epsilon(\mu+H_{eff})+\sqrt{\epsilon
U/2}(i\alpha^a_k+\gamma^a_k)\right]
+\prod_k\left[1+\epsilon(\mu-H_{eff})+\sqrt{\epsilon
U/2}\hspace{.1cm}(i\alpha^a_k-\gamma^a_k)\right]
\nonumber\\
&+&\prod_k\left[(1+\epsilon\mu+\sqrt{\epsilon U/2}\hspace{.1cm}i\alpha^a_k)^2
-(\epsilon H_{eff}+\sqrt{\epsilon
U/2}\hspace{.1cm}\gamma^a_k)^2\right])
\label{detsol}
\end{eqnarray}
and extract the contributions which survive in the continuum time limit,
expressed by $\epsilon\rightarrow 0, M\rightarrow\infty, M\epsilon=\beta$.
The integration over the Hubbard decoupling fields can be executed exactly
which yields in the continuum time limit
\begin{equation}
\int^G_{\alpha,\gamma} det B
=2e^{\beta\mu}\left[cosh(\beta(\mu-\frac{U}{2}))e^{-\beta U/2}+cosh(\beta
H_{eff}(y^a,\{z_r^{(b_r)}\})\right]
\label{Eq:detB}
\end{equation}
Using this result together with Eq.\ref{def:H} and Eq.\ref{avZ2}
in the expression for the free energy per site (respectively the thermodynamic
potential) $F=
-T\lim_{n\rightarrow0}(\langle Z^{(n)}\rangle-1)/n$ one finds (after
one more integration over the $y^a$-fields)
\begin{eqnarray}
\beta F=\frac14\bar{\chi}^2+\frac{\beta}{2}(q_1-1)\bar{\chi}
&+&\frac14\beta^2\sum_{r=1}^{\infty}m_r(q_{r}^2-q_{r+1}^2)
-\beta\mu-ln2
\label{Eq:f}\\
& &-\lim_{r\rightarrow\infty}\frac{1}{m_{r}}\int_{z_{r+1}}^G
ln(\left[\int_{z_{r}}^G\{...\{\int_{z_2}^G\{\int_{z_1}^G
{\cal{C}}(\{z_r\})^{m_1}\}^{m_2/m_1}\}^{m_3/m_2}
...\}^{m_{r}/m_{r-1}}\right],\nonumber\\
& &\hspace{-4cm}{\cal{C}}(\{z_r\})\equiv cosh(\beta H_{eff}(0,\{z_r\}))+
e^{-\beta(U+\bar{\chi})/2}cosh(\beta(\mu-U/2))
\label{Eq:C}
\end{eqnarray}
where $\bar{\chi}=\beta(\tilde{q}-q_1)$ is the so-called single-valley susceptibility.
The problem is now solved up to integrations over the Gaussian $z$-spin field
distributions;
in other words, we have reduced the thermodynamics of the given model to that
of the standard $SK$-model, which is contained as the special case $\mu=i\pi
T/2,U=0$, with a different kernel ${\cal{C}}$. The field-independent part of
${\cal{C}}$ is responsible for the competition between Hubbard coupling,
chemical potential, and the self-consistent single valley susceptibility, which
must be determined self-consistently, is obvious.
The $RSB$-effects are notoriously difficult to treat. They are of importance for the
low $T$ thermodynamic behaviour, and found to be even more crucial for the
quantum-dynamical fermionic correlations. These are derived from the generating
functional below. The model is particle hole symmetric ($PHS$) at $\mu=U/2$, where
the spin glass temperature is maximal; under $PHS$ Eq.\ref{Eq:f} still shows a
$U$-dependence which is necessary to explain for example the magnetic breakdown for
sufficiently strong attractive Hubbard interaction (negative $U$ and for arbitrary
$\mu$).
%
\subsection{Generating functional for quantum-dynamical fermion correlations}
\label{subsec:genfunct}
The imaginary time single-fermion propagator (and N-fermion propagators too)
\begin{equation}
{\cal{G}}^{(a)}_{ij,\sigma}(\tau_k,\tau_{k^{\prime}})=
-\langle \psi_{i\sigma}^a(\tau_k)\bar{\psi}_{j\sigma}^a(\tau_{k^{\prime}})\rangle
=-\delta_{ij}\langle\psi_{i\sigma k}^a\bar{\psi}^a_{i\sigma
k^{\prime}}\rangle
\label{Eq:G-def}
\end{equation}
can be derived by derivatives with respect to the generating fields $\eta$
from the generating functional
\begin{equation}
\Phi(\eta,\bar{\eta})=ln\left[\prod\int {\cal{D}}\Psi
e^{-\bar{\Psi}B\Psi+\eta\bar{\Psi}-\bar{\eta}\Psi}\right]
\end{equation}
Let us symbolize Gaussian integrals over arbitrary decoupling-fields by
$\prod\int_{_{\{X\}}}^G$, where $\{X\}$ stands either
for $\{z\}$-spin or $\{\alpha,\gamma\}$-Hubbard fields, and by $\int{\cal{D}}\Psi$
for the Grassmann field integration. Using
\begin{equation}
\int{\cal{D}}\Psi
e^{-\bar{\psi}B\Psi+\eta\bar{\Psi}-\bar{\eta}\Psi}=\det B e^{\bar{\eta} B^{-1}\eta}
\end{equation}
we calculated the inverse matrix $B^{-1}$ in order to obtain the propagator.
We choose
$\tau>\tau^{\prime}$ with $\tau\equiv\tau_{k_i}, \tau^{\prime}\equiv\tau_{k_f}$
and obtain the (disorder-averaged, local) fermion propagator as
\begin{eqnarray}
& &{\cal{G}}^{(a)}_{\sigma}(\tau^{\prime}-\tau)=
-\lim_{M\rightarrow0}
\left[(\prod_r\int_{\{z_r\}}^G)
\prod_{a'}\int_{y^{a'}}^G\int_{\alpha_k^{a'}}\int_{\gamma_k^{a'}}
\right]
\prod_{k=k_i}^{k_f-1}\left[1+\epsilon\mu+\epsilon\sigma
H_{eff}(y^a,\{z\})+\sqrt{\frac{\epsilon
U}{2}}(i\alpha^a_k+\sigma\gamma^a_k)\right]\nonumber\\
& &\left[1+\prod_{k=0}^{M-1}(1+\epsilon\mu-\epsilon\sigma
H_{eff}(y^a,\{z\})+\sqrt{\frac{\epsilon
U}{2}}(i\alpha^a_k-\sigma\gamma^a_k))\right]/{\cal{N}},
\end{eqnarray}
where
${\cal{N}}=\left[\prod\int_z\prod\int_y\int_{\alpha}\int_{\gamma}\right]det B$.
Splitting the product of the second bracket into one over the interval
$k_i,k_f$ and one for the rest, the resulting form allows to solve the
Gaussian Hubbard-integrals. Terms of order $\epsilon^{3/2}$ do
not survive the continuum limit and the exact solution becomes
\begin{equation}
{\cal{G}}^{(a)}_{\sigma}(\tau)=-(\prod_r\int_{\{z_r\}}^G)\prod_{a'}\int_{y^{a'}
}^G(
e^{\mu+\sigma H_{eff})\tau}+e^{\beta(\mu-\sigma H_{eff})}
e^{(\mu+\sigma H_{eff}-U)\tau}/{\cal{N}}
\end{equation}
After Fourier transformation from imaginary times to imaginary frequencies
($\epsilon_n\equiv (2n+1)\pi T$) the fermion Green function in $1RSB$,
which illustrates $RSB$-effects on the propagator in a relatively simple form,
is obtained as
\begin{equation}
\hspace{-.15cm}{\cal{G}}^{1RSB}_{\sigma}(\epsilon_n)=\int_{z_2}^G
[\int_{z'}^G \bar{C}(z',z_2)^m]^{-1}
\int_{z_1}^G \bar{C}(z_1,z_2)^{m-1}\int_y^G
\left[
\frac{e^{\beta(\mu+\sigma H_{eff})}+1}{i\epsilon_n+\mu+\sigma H_{eff}}
+\frac{e^{\beta(\mu-\sigma H_{eff})}(e^{\beta(\mu-U+\sigma H_{eff})}+1)}
{i\epsilon_n+\mu-U+\sigma H_{eff}}
\right]
\label{Eq:G-result}
\end{equation}
with normalizing factor $\bar{C}(z_1,z_2)\equiv 2
e^{\beta(\mu+\chi/2)}{\cal{C}}(z_1,z_2)$ and ${\cal{C}}$ given by Eq.\ref{Eq:C}.
The result reduces to the known limits: in case of vanishing effective field
$H_{eff}=0$ the propagator of the Hubbard model's atomic limit is retrieved
\cite{kotliar.et.al} and, for $U=0$, the fermionic Ising spin glass limit is
retained \cite{rohf}. The result Eq.\ref{Eq:G-result} displays the $U$-correlation
induced two-level splitting, which is broadened into bands by the
statistically distributed spin glass fields $H_{eff}(z$).
This smearing of delta-peaks into bands might lead to less singular expansions
in powers of hopping terms than those expansions encountered in the pure
Hubbard model.
We shall further solve below the y-integration for all $T$ and at $T=0$ also the
$z_1$-integration in order to get a most explicit analytic form.
The remaining numerical task is still not quite easy, since all parameters on
which the effective field depends and the Parisi block parameter $m=m(\mu,U,T)$
must first be found as selfconsistent solutions and then used in the calculation of
the density of states. As Eq.\ref{Eq:G-result} suggests the procedure can be
extended to formulate $N$-particle propagators.
\section{Dependence of freezing temperature on repulsive and on
attractive Hubbard couplings $U$}
\label{section:Tf}
\begin{figure}
\centerline{
\epsfxsize=12cm
\epsfbox{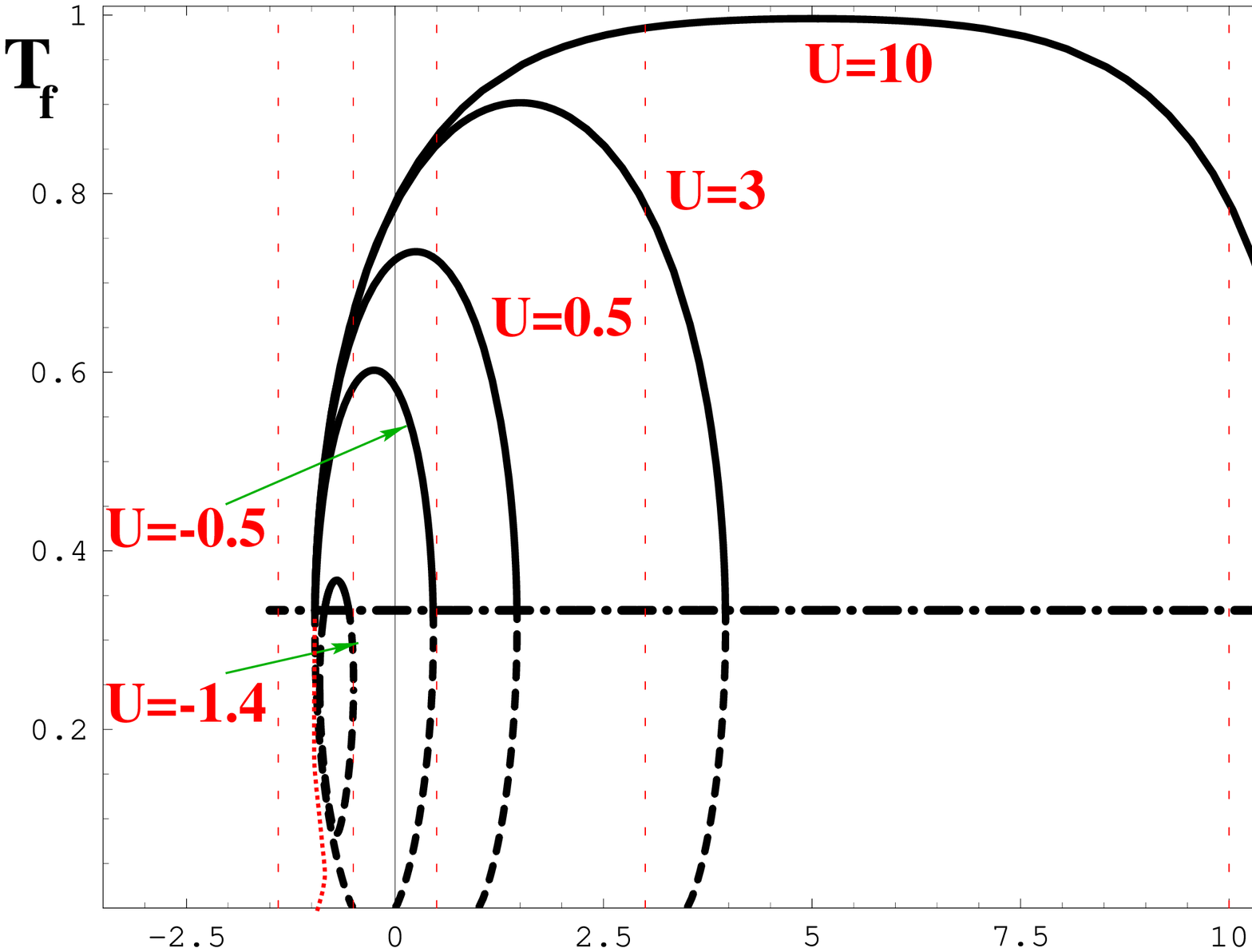} }
\caption{Phase boundaries of the solutions which extremize the free energy
(\ref{Eq:Tf}), shown as a function of chemical potential $\mu$ and
for a set of fixed Hubbard couplings $U = -1.4, -0.5, 0.5, 3$, and
$10$ (in units of $J$). Solid bold lines are solutions of
Eq.(\ref{Eq:Tf}) in the regime of continuous transitions. Dashed
bold lines, beneath the dash-dotted bold tricritical line at
$T_{f3}=1/3$, are solutions of Eq.(\ref{Eq:Tf}) too but describe
paramagnetic only stability limits preempted by first order
transitions. First order transition lines, which depend on $RSB$,
are shown for $1RSB$ for $U=10$. Dashed vertical lines at $\mu=U$
show (as a guide for the eye) how far $\mu$ can exceed the Hubbard
coupling $U$ (which is also the half-filling limit at $T=0$)
before spin glass order breaks down. Each curve is symmetric with
respect to its particle-hole symmetric point at $\mu=U/2$. }
\label{Tf}
\end{figure}

%
The temperature $T_f$ at which spin glass freezing occurs depends on the chemical
potential $\mu$ and on the Hubbard coupling $U$.
The reduction of doubly occupied sites under increasing $U>0$ favors magnetic order
and thus counteracts the effect of a growing deviation from particle hole symmetry,
described by the parameter $\Delta_{PHS}\equiv|\mu-\frac12 U|$.
Thus the freezing temperature tends towards the SK-value in the infinite
repulsion limit for fixed $\Delta_{PHS}$.
On the other hand, negative (attractive) Hubbard interactions act in the same
direction as 'particle pressure' from the particle reservoir:
both enhance double occupation.
At a critical negative $U$ the magnetic phase will break down.\\
The derivation of the nonperturbative selfconsistent $T_f$-equation follows the
standard procedure for the fermionic generalization of the
$SK$-model.
The results of this section identify the regimes of continuous and of
discontinuous transitions below a tricritical temperature called $T_{f3}$.\\
The selfconsistent equation for the freezing temperature of the infinite range model
reads (in units of $J$)
\begin{equation}
T_f=1/(1+e^{-(1+U)/(2T_f)}cosh((\mu-U/2)/T_f)),
\label{Eq:Tf}
\end{equation}
which reduces in the special case $U=0$ and $\mu=i\pi T/2$ to the standard
SK-model result. \\
The numerical evaluation of $T_f$ for several fixed values of $U$,
including attractive interactions, is displayed in Figure \ref{Tf}.
The Figure illustrates the competition between chemical potential and
Hubbard repulsion.\\
The vertical dashed lines at $\mu=U$ drawn in Fig.\ref{Tf} will be found
later to represent the exact limit of half-filling at $T=0$, while
particle hole symmetry (not shown in the Figure) lies at $\mu=\frac12 U$.
Figure \ref{Tf} shows specific cuts through the model's multidimensional phase
diagram which provide here a complete picture of its dependence on chemical potential
and Hubbard interaction. Only the first order thermodynamic transition lines
$T_{f1}(\mu)$ (defined by the crossing of magnetic and nonmagnetic free
energies) all resemble the one shown around $\mu\approx 10.9$ and depend on $RSB$.
They are not yet better determined than in $1RSB$. The back-swing observed for
$T_{f1}\rightarrow0$ may well occur at the crossover temperature
where the next higher $RSB$-step is required. The full
$RSB$-solution is not yet known, but is expected to lie closer to
the stability limit. Thus Figure \ref{Tf} determines the phase diagram's
$\mu$-dependence apart from small $RSB$-corrections near $T_{f1}=0$.
In the large $U$-limit, the $T_f$-curve approaches the shape of a box
with height $1$ and a width given by the value of the chemical
potential at the tricritical point. This is obtained in
Eq.\ref{Eq.:tricritical} below.
%
\subsection{The tricritical temperature $T_{f3}$ and corresponding chemical
potential $\mu_{_{f3}}(U)$}
%
The tricritical temperature, included in Figure \ref{Tf}, appears as the lower limit
of second order freezing transitions and remains $\frac13$ independent of $U$.
This is readily found by solving Eq.\ref{Eq:Tf} together with the condition
$cosh((\mu-U/2)/T_{f3})=2 \exp((1+U)/(2T_{f3}))$, where the
$O(q^2)$-term of the order parameter equation changes sign. For deviations from
particle hole symmetry (half-filling) which are larger than a tricritical value
$|\mu_{_{f3}}-U/2|$ given in Eq.(\ref{Eq.:tricritical}), the order parameter
can no more vanish continuously.\\
A special feature is the breakdown of spin glass order due to sufficiently
strong attractive (negative) $U$ couplings.\\
As Figure \ref{Tf} shows, minimum values in the range of negative
Hubbard couplings $U_{min}$ and of the chemical potential exist, below
which a spin glass cannot exist.
The tricritical chemical potentials are symmetrical with respect
to $PHS$ at $\mu=\frac{U}{2}$ and one finds (recalling that $\mu, T, U$ are
given in units of $J$)
\begin{equation}
\{T_{_{f3}},\mu_{_{f3}}(U)-\frac{U}{2}\}=\{\frac13,\hspace{.2cm}\pm\frac13
arccosh(2e^{3(1+U)/2})\}
\label{Eq.:tricritical}
\end{equation}
For homogeneous attractive Hubbard interaction
\begin{equation}
U > U_{0} = -1-\frac23 ln 2
\end{equation}
limits the regime of continuous transitions. In a very small range $U<U_{0}$
discontinuous transitions still exist.
\subsection{Implications of magnetic breakdown for large negative $U<U_{min}$}
The destruction of magnetic order by sufficiently attractive Hubbard interaction
is due to a redistribution of density of states.
A critically high density of paired states breaks down the magnetic order and
in principle would create a coherent superconducting state in cases
when fermion hopping- or pair-hopping mechanisms like those given in Eq.\ref{hopping}
are present. The nonlocal magnetic interaction $J_{ij}$ by itself can however not
generate a coherent superconducting state.
\\
As discussed in a chapter \ref{sec:randomU}, $J$-$U$ models with
additional random $U$ and in particular with ($U_A>0,U_B<0$)
include extreme proximity- (vicinity-) effects between local pairs and
local moments on neighboring sites. It will be very interesting to study the
instability of such phases towards delocalization by Hamiltonians of the
type \ref{hopping}a) and b).
%
\section{Band- and gap-structures in the presence of frustrated magnetic order}
\label{section:bands}
%
\subsubsection{The fermionic density of states of the SK model with
Hubbard interaction in replica symmetric approximation}
%
In models as hard to solve as Hubbard- and spin glass models the density
of states has the virtue of being one of the simpler quantities with yet
a broad physical significance. It is important for many one particle properties.
Thermodynamic behaviour as observed in quantities like particle number
(as a function of chemical potential), internal energy, specific heat, thermopower
etc. depend on it. Detailed properties of the $DOS$ near the Fermi level
influence the stability of the insulating phase against fermion hopping processes
in itinerant model extensions. Due to the Ward identity for charge conservation
\cite{ro.epl} one finds that the $DOS$ also determines the amplitude of the diffusive
two-particle propagator. The $DOS$ can predict the existence of a metal-insulator
transition by means of the filling of an energy gap at the Fermi level;
however it is insufficient to decide where and under which conditions exactly such
a transition takes place, since this requires to calculate the decay of the diffusion
constant by means of a two-particle correlation function. It is one goal of the
present paper to construct the basis for such calculations.\\
The replica symmetric approximation is known to be unstable everywhere in the
ordered phase (in fermionic extensions as well as in standard $SK$-models).
While $RSB$-corrections remain small only down to characteristic crossover
temperatures, some crude features like the splitting of bands remain correct even
down to zero temperature. It provides thus a simple and partially meaningful basis
for improved calculations. In particular, under its simplifications in the zero
temperature limit, the analytical expression for the Green function's
spectral weight $\rho(\epsilon)=-\frac1\pi Im\{G^R(\epsilon)\}$
is simple and reads in terms of energy variable $E=\epsilon+\mu$ as
\begin{eqnarray}
\rho_{_0}(E)&\equiv&\rho(E)|_{_{0RSB}}=\frac{1}{\sqrt{2\pi q}}
e^{-(E+\bar{\chi})^2/(2q)}\theta(min[-\bar{\chi},U-\mu-\bar{\chi}/2]-E)\\
&+&e^{-(E-U)^2/(2q)}\theta(E-(2U-\mu+\bar{\chi}/2))\theta(\mu-\bar{\chi}/2-E)
+e^{-(E-U-\bar{\chi}/2)^2/(2q)}\theta(E-max[U+\bar{\chi},\mu+\bar{\chi}/2]),
\label{dosT0}\nonumber
\end{eqnarray}
where the parameters must be understood as $q\equiv q(T=0,\mu,U)|_{_{0RSB}}$,
$\bar{\chi}\equiv \bar{\chi}(T=0,\mu,U)|_{_{0RSB}}=\chi(T=0,\mu,U)|_{_{0RSB}}$.
These parameters $q$ and $\bar{\chi}$ are determined by the zero temperature
limit of their selfconsistency equations given by
\begin{equation}
q=1-|\nu-1|,\quad\quad
\nu-1=erf(\frac{\mu-U-\bar{\chi}/2}{\sqrt{2q}})\theta(\mu-\frac{\bar{\chi}}{2}-U)
,\quad\quad \bar{\chi}=\sqrt{\frac{2}{\pi
q}}e^{(\mu-U-\bar{\chi}/2)^2/(2q)}\theta(\mu-\bar{\chi}/2-U)
\end{equation}
Eq.\ref{dosT0} correctly predicts the crude band structure:
\begin{eqnarray}
& &If\hspace{.2cm}U<\mu-\bar{\chi}/2\hspace{.2cm}:\rightarrow{\rm three\hspace{.15cm}
bands\hspace{.2cm}for}\hspace{.2cm} E<U-\mu-\bar{\chi}/2,
\hspace{.2cm}2U-\mu+\bar{\chi}/2<E<\mu-\bar{\chi}/2,\hspace{.2cm}E>\mu+\bar{\chi}/2
\nonumber\\
& &If\hspace{.2cm}U>\mu-\bar{\chi}/2\hspace{.2cm}:\rightarrow {\rm two\hspace{.15cm}
bands\hspace{.2cm} for}\hspace{.2cm}E<-\bar{\chi},\hspace{.2cm}E>U+\bar{\chi}\nonumber
\label{bandstructure}
\end{eqnarray}
The values of these band limits (at $T=0$) will change under
improvement of the approximation. That means, with each step of
replica symmetry breaking the single valley susceptibility
$\bar{\chi}$ will decrease further towards zero and thus deviate more and more from
the equilibrium susceptibility. This turns the band limits of the lower
approximations into crossover lines of the higher ones. Finally these crossover lines
will lie dense giving way to a totally softened decay of the density of states
towards its pseudogap (or Hubbard gap, respectively).
In order to avoid misunderstandings we remark that the Gaussian
bands given in Eq.(\ref{dosT0}) will not shrink into delta peaks
when the order parameter vanishes at and above $T_f$.
The random interaction maintains the Gaussian broadening above $T_f$.\\
The full Green function (in the finite temperature technique) is given by the
spectral representation
${\cal{G}}(\epsilon_l)=\int_{-\infty}^{\infty}d\epsilon
\frac{\rho(\epsilon,T)}{i\epsilon_l-\epsilon}$
where $\epsilon_l=(2l+1)\pi T$
and the finite $T$ replica symmetric $DOS$ $\rho_{_0}(E,T)$ is
given by
\begin{eqnarray}
\rho_{_0}(T,E)&=&\frac{e^{-\frac{1}{2}\bar{\chi}/T}}{4\pi\sqrt{T\bar{\chi}}}
((e^{-E/T} + e^{-\mu/T})
\int_{-\infty}^{\infty}dz
e^{-\frac{1}{2}z^2}e^{-(\sqrt{q}z+E)^2/(2T\bar{\chi})}\xi^{-1}(\mu,U,T,z)\nonumber\\
&+&(e^{(\mu-U)/T}+e^{(E-U)/T})
\int_{-\infty}^{\infty}dz
e^{-\frac{1}{2}z^2}e^{-(\sqrt{q}z+E-U)^2/(2T\bar{\chi})}\xi^{-1}(\mu,U,T,z))
\end{eqnarray}
with
\begin{equation}
\xi(\mu,U,T,z):=e^{-(U+\bar{\chi})/(2T)}cosh(\frac{\mu-U/2}{T})
+cosh(\frac{\sqrt{q}\hspace{.1cm}z}{T})
\label{dosfiniteT}
\end{equation}
\subsubsection{Fermion concentration}
\begin{figure}
\centerline{
\epsfxsize8cm
\epsfbox{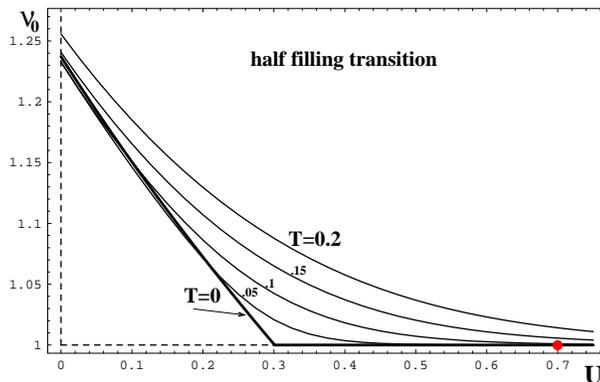}
}
\caption{Thermal smearing of the sharp half filling transition at $T=0$ as shown
by the fermion concentration $\nu_0(T)$ for $\mu=0.7$ in
replica symmetric approximation.
Temperatures are given in units of $J$. The critical Hubbard coupling
$U_{c}=\mu-\bar{\chi}(\mu,U_c)/2$ in $0RSB$-approximation and its exact position (dot)
at $\mu=U$ for vanishing nonequilibrium susceptibility $\bar{\chi}$ (see below).}
\label{fillingarray}
\end{figure}

The fermion concentration for given chemical potential is calculated using
Eq.(\ref{dosfiniteT}) in
\begin{equation}
\nu(\mu,U,T)=\int_{-\infty}^{\infty}dE \frac{1}{e^{(E-\mu)/T}+1}\rho(E)
\label{Eq:nu-rs}
\end{equation}
The $RS$-approximation of Eqn.\ref{Eq:nu-rs} already predicts correctly the
existence of a continuous half-filling transition. The half-filling transition
is defined as a transition from filling $\nu=1$ to $\nu\neq1$.
Figure \ref{fillingarray} displays its thermal smearing for $\mu=0.7$.
The critical point at $U_c=\mu-\bar{\chi}(\mu,U_c)/2$ will be strongly corrected
by $RSB$-effects (as derived in the following sections). Since the
single-valley susceptibility decreases with $RSB$ and must be
expected to vanish in $\infty RSB$ the half-filling transition
occurs at $U_c=\mu(=0.7$ in this case).
One aspect of $RSB$ and of the unbroken approximation becomes
obvious from Fig.\ref{fillingarray}: temperatures above $0.15[J]$
are enough to render $RSB$-effects small enough and results
obtained in $0RSB$ free from obvious failure; the effects become
large near $T=0$ and better approximations are required. Yet the
$0RSB$-approximation has the virtue to predict correctly the
existence of a transition and to provide crude analytical
solutions (without too many integrals).\\ Moreover, $RSB$ has the
power to generate nontrivial critical behaviour in the range-free
model; mean field critical behaviour only refers to the
elimination of critical fluctuations in real space. We shall see
this further below from the $4RSB$-results of section
\ref{section:4RSB}.\\ We conclude this section with a
demonstration of the typical difference between band asymmetries
generated by ferromagnetic order (assuming here a model with
nonvanishing ferromagnetic mean interaction $\langle J\rangle>0$
and $U=0$, Figure\ref{Fig:band-asymmetry} (left)) and the effects
of the Hubbard interaction with $U>0$ while $\langle J\rangle=0$
in Fig.\ref{Fig:band-asymmetry} (right). We chose a strong doping
case with a pronounced central band. The spin glass gaps are
present in both cases. Ferromagnetic order shifts spectral weight
from the upper magnetic band into the lower one and simultaneously
renders the nonmagnetic central band asymmetric as well. On the
right side the asymmetry above freezing is only due to the
deviation from half-filling, which below $T_f$ becomes accentuated
by the spin glass gap at $\mu$, which splits off the central band
from the upper magnetic band.\\

Figure 3 appended (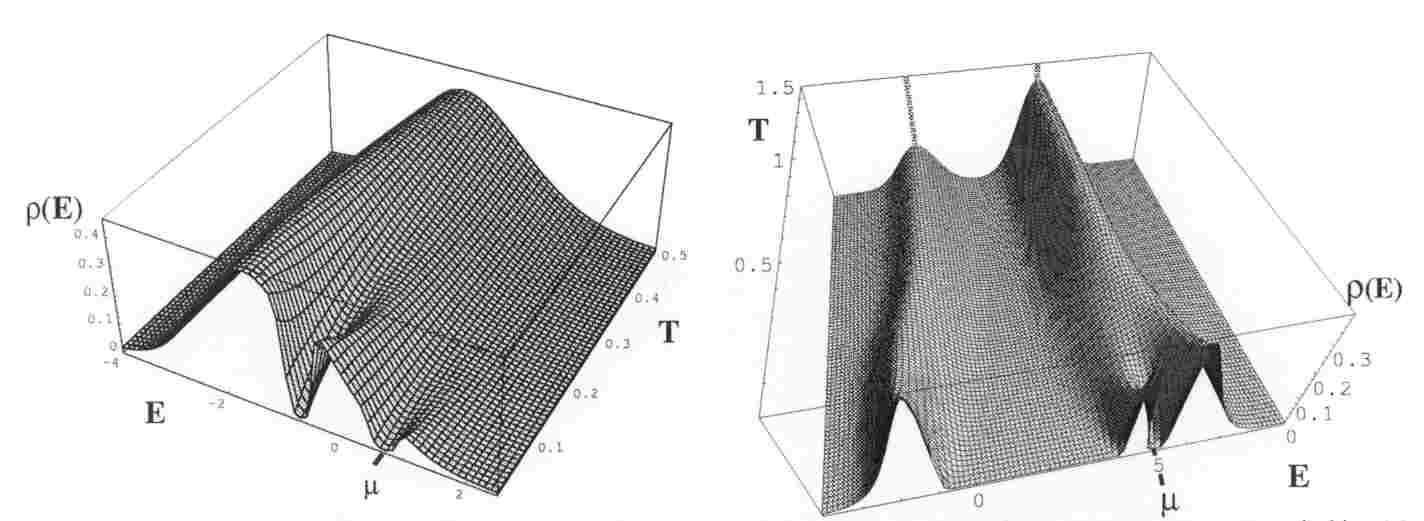) at e.o.f.
\begin{figure}
\caption{
Comparison of strong band asymmetry effects induced either by partially ferromagnetic
interaction (left) with $\langle J_{ij}\rangle=1.5, \mu=0.7, U=0$ and selfconsistent
ferromagnetic order or (right Figure) due to a large chemical potential $\mu=5$
with Hubbard coupling $U=4.4$ and $\langle J_{ij}\rangle=0$ above half filling.
In the latter case the lower spin glass gap is immersed in a large Hubbard gap
and the upper spin glass gap hosts the Fermi level, while in the left Figure
only spin glass gaps, symmetrically placed with respect to $E=0$,
are excavated in the presence of strong additional ferromagnetic order.
The high temperature regime of the right hand side Figure shows two energy levels
with Hubbard splitting in the clean atomic limit; the random magnetic interaction
with variance $J$ (set equal to $1$) distributes their weight into bands which
remain well separated for the chosen ratio $U=4.4$.
}
\label{Fig:band-asymmetry}
\end{figure}
\section{Solutions with 1-step replica symmetry breaking}
\label{section:1RSB}
%
Since $RSB$-effects are strongest in the low temperature regime, we show first
the $1RSB$ results of all relevant physical parameters of our present $J-U$ model (1)
in the range between $T=0$ and moderate temperatures $T\approx0.25(J)$.
These are collected in Fig.\ref{fig:1rsb-vs}.
For $T=0$ the self-consistency equations show that the
$(T=0,U=0)$-solution can be used to find the $(T=0,U>0)$ solutions simply
by applying the shift $\mu\rightarrow\mu+U$. This relation can be
concluded from the exponential divergence $\sim e^{-\beta(\mu-U+\bar{\chi}/2)}$
(in case that $\mu>U/2$) of expression (17) as $\beta\rightarrow\infty$ and,
of course, is confirmed by our numerical evaluations.\\

Figure 4 (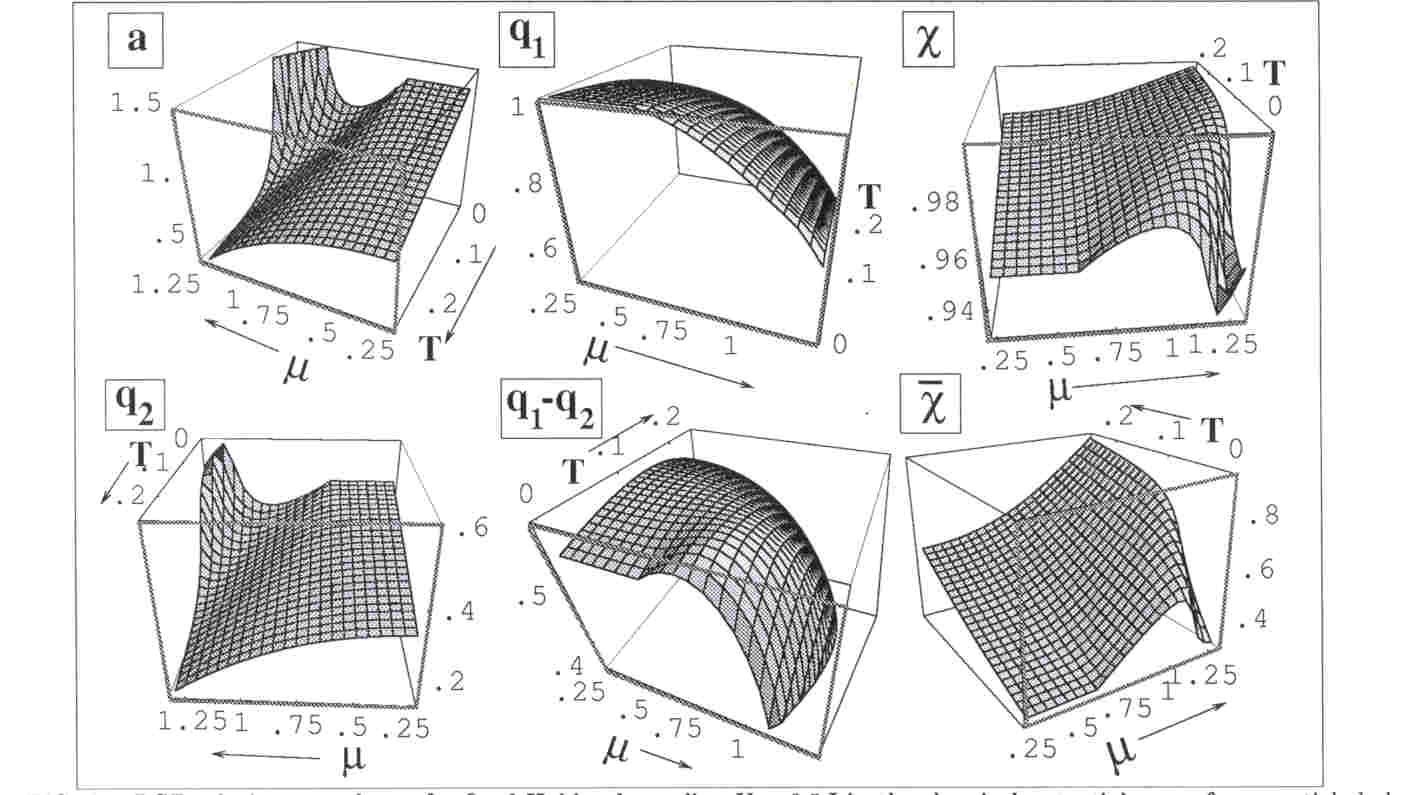) appended at e.o.f.
\begin{figure}
\caption{$1RSB$-solutions are shown for fixed Hubbard-coupling $U=0.5 J$ in the
chemical potential range from particle-hole symmetry, $\mu=U/2$, to almost the
discontinuous breakdown of the magnetic phase at $\mu\approx 1.4$, and in the
temperature-interval $0\leq T \leq \frac14$: the $T$-normalized Parisi parameter
$a(\mu,T)=m(\mu,T)/T$, the order parameters $q_1(\mu,T)$, $q_2(\mu,T)$ and their
difference, the equilibrium susceptibility $\chi(\mu,T)$ and single valley
susceptibility $\bar{\chi}(\mu,T)$. Traces of the nonanalytic behaviour caused by
the half-filling transition at $(\mu=U+\bar{\chi}/2,T=0)$
disappear with increasing temperature.}
\label{fig:1rsb-vs}
\end{figure}
Using these solutions we obtain the density of states. Characteristic cases are
displayed in Fig.\ref{HSKdos}. On the left $\rho(E,T)$ for $\mu=1, U=0.5$
demonstrates for large filling how the broader Hubbard-SG gap and the upper SG-gap
are filled by thermal excitations. The right Figure $\rho(E,\mu)$ shows a
clear cut nonmagnetic central band together with a visible combination of gap widths
at the very low temperature $T=0.01$ and $U=\frac12$.\\

Figure 5 (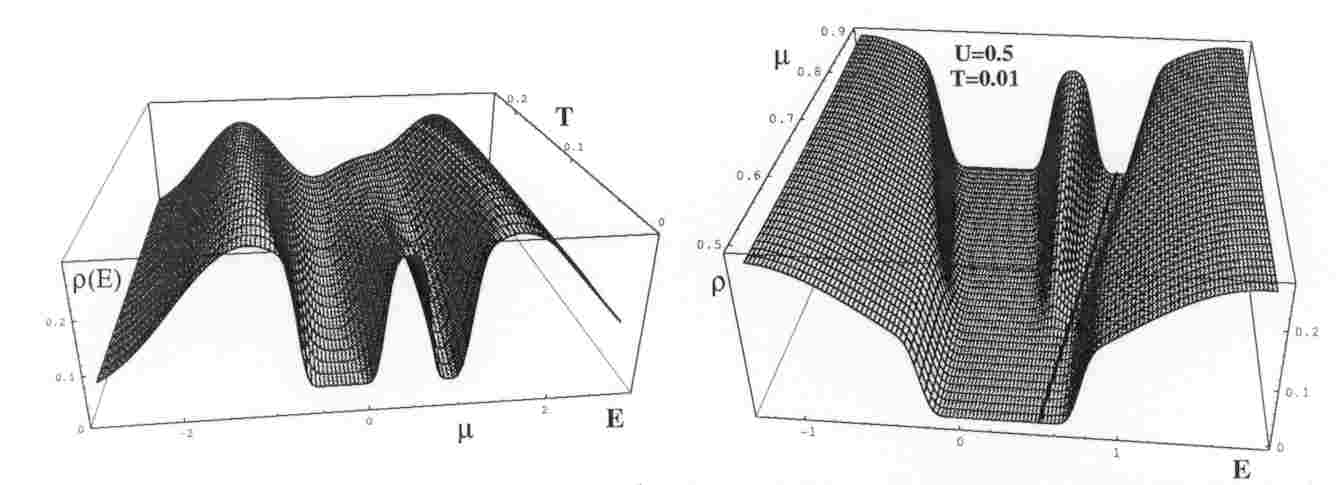) appended at e.o.f.
\begin{figure}
\caption{Density of states $\rho(E=\epsilon+\mu,T)$ for $\mu=1,U=\frac12$
(left Figure) showing thermal smearing of Hubbard- and spin glass gaps and
(in the right Figure) weakly thermalized gaps in $\rho(E,\mu)$ for $T=0.01,U=\frac12$
(all energies in units of $J$). The position of the Fermi level is
indicated by a thick line, showing the crossover from a spin glass gap
into the combined Hubbard-and spin glass gap.
Both figures are obtained in $1RSB$-approximation.}
\label{HSKdos}
\end{figure}
The symmetry between lower and upper magnetic band demonstrate that the half-filling
transition coincides with the disappearance of the central band as the chemical
potential falls below the critical value. As a guide to the eye a thick line shows
the position of the Fermi level, which leaves the SG-gap below a critical
$\mu(U)$ and moves deeper into the Hubbard regime as $U/\mu$ further increases.
The presence of a central nonmagnetic band is synonymous with a deviation
from half filling in this model. It exists for chemical potentials within the
interval $\mu_k<\mu<\mu_c$, where $\mu_k\rightarrow0$ for $k=\infty$ steps
of $RSB$ and $\mu_c$ denotes the threshold value for the spin glass phase.
%
\section{Half filling phase transition at $T=0$: effect of the
competition between Hubbard repulsion and chemical potential}
\label{section:half-filling-transition}
%
\subsection{Splitting and recombination of gaps}
The half-filling transition is sharply defined at $T=0$. We summarize the
superposition and splitting of Hubbard- and spin glass gap(s) in the following
Figure \ref{gaps}.
The spin glass gap (finite in 0$RSB$) turns into a pseudogap at the Fermi level in the
exact solution, since $RSB$ fills spectral weight into the bright/vacant regions
exempting only the Hubbard gap (blue) and the pseudogap line along $E=E_F$.
Thus Figure \ref{gaps} exhibits a change between two very different insulating phases:
in the first phase, for $\mu>U$, the Hubbard interaction has effects only remote from
the Fermi level, while in the half-filled regime ($\mu<U$) the Fermi level changes
over into the Hubbard gap regime.
\begin{figure}
\centerline{
\epsfxsize15.cm
\epsfbox{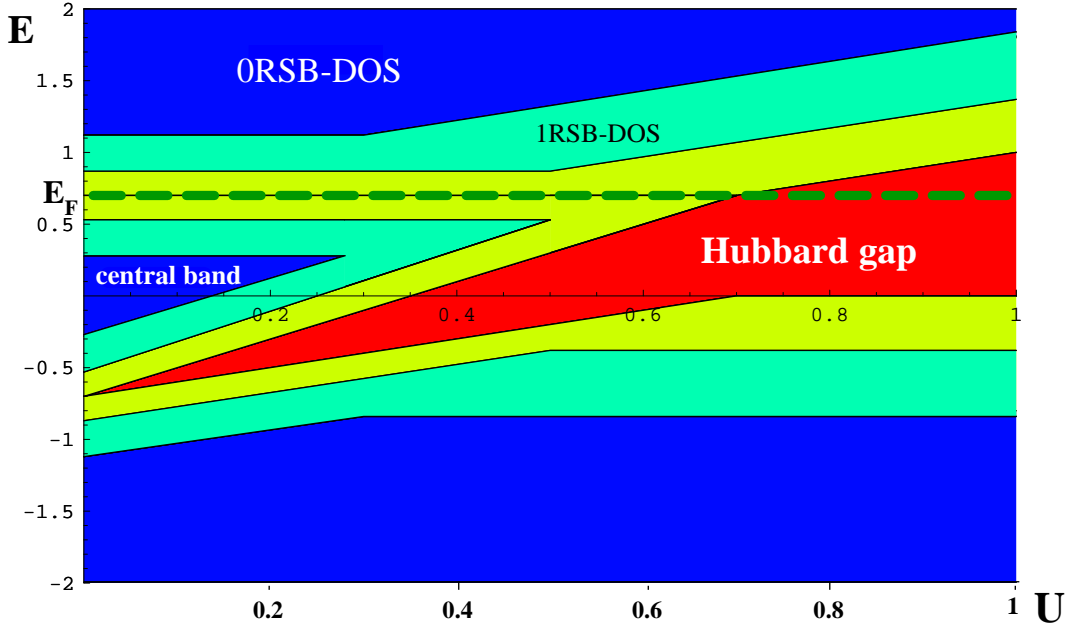}}
\caption{
This Figure illustrates the crossover (along the dashed line) of the Fermi level
between different gaps in the $E,U$-plane at $T=0$. Dark grey regions show
nonvanishing density of states in 0$RSB$-approximation, light grey the extension
of these domains in $1RSB$.
The $DOS$-landscape is viewed from above. For a chosen $E_F=\mu(=0.7$ in the
example shown), $E_F$ pins the
upper spin glass gap for $U<\mu$ and moves into the Hubbard gap for $U>\mu$.
At $T=0$ the central nonmagnetic band becomes fully suppressed for any Hubbard
repulsion stronger than $U=\mu-{\bar{\chi}}/2$, which then implies half filling.
In the exact solution the gap region is reduced and consists of Hubbard gap
and spin glass pseudogap (of zero width).}
\label{gaps}
\end{figure}
The way in which $2RSB$ and higher corrections fill vacant regions
of Figure \ref{gaps} with small but finite spectral weight becomes obvious from the
results of our paper. Here we want to avoid information overflow in the Figure.
For the same reason, small deviations from straight lines are not shown,
which occur beyond half filling (but, as found below in the paper) are identified as
$RSB$-artifacts. The exact solution finally realizes $\rho(E)=0$ only along the
straight line at $E_F$ and in the Hubbard gap region. The transition between two
different types of insulating phases remains as described by Figure
\ref{gaps}.\\
Another type of soft filling into the regions, where Figure \ref{gaps} still
shows evacuated spectral weight, occurs due to thermal effects.
The transition between Hubbard/spin glass gap phase into the pure
spin glass gap phase will be manifested for example in the specific heat (as
usual specific heat results are reasonable down to temperatures of the order
$T>T_k\approx E_{gk}$, where $E_{gk}$ denotes the gap energy in $kRSB$).\\
In which way transport processes will scatter weight into the gap regimes is a
matter of speculation for the moment. Essential parts of the present paper are
however concerned with a sufficiently good determination of the
fermion propagators for the insulating phase, which then provide the means to
analyze transport effects in a perturbation theory.
\section{Dependence of order parameter, susceptibilities, and filling factor on
the chemical potential: a 2$RSB$-analysis at zero temperature
}
\label{section:2RSB}
%
The half filling transition requires a more refined analysis of
replica symmetry breaking effects. The dependence on the chemical
potential is therefore calculated in a selfconsistent $2RSB$-approximation for the
full range of $\mu$ for which the spin glass phase exists. We
determine all relevant parameters in this section, distinguish
$RSB$-artifacts from true $\mu$-dependence, and in particular determine the
$\nu(\mu)$ which helps to change arbitrarily between dependence on
filling (fermion concentration) or on chemical potential. The
filling factor is also linked to the integrated density of states
(provides internal consistency check), which is determined in
greater detail later in the paper.
\subsection{Analytical input for the numerical solution at $T=0$}
\begin{figure}
\centerline
{
\epsfxsize8.8cm
\epsfbox{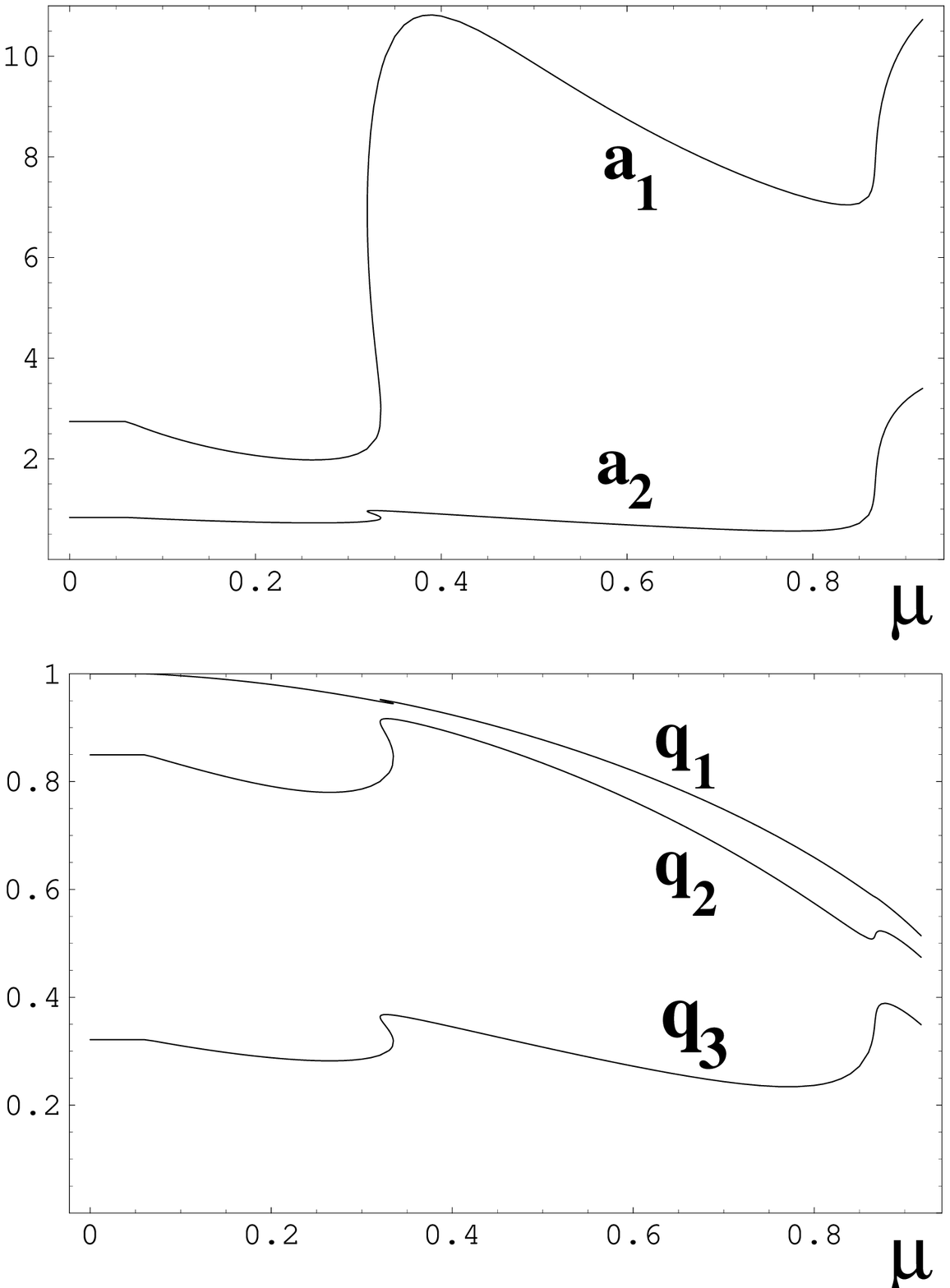}
\epsfxsize9.5cm
\epsfbox{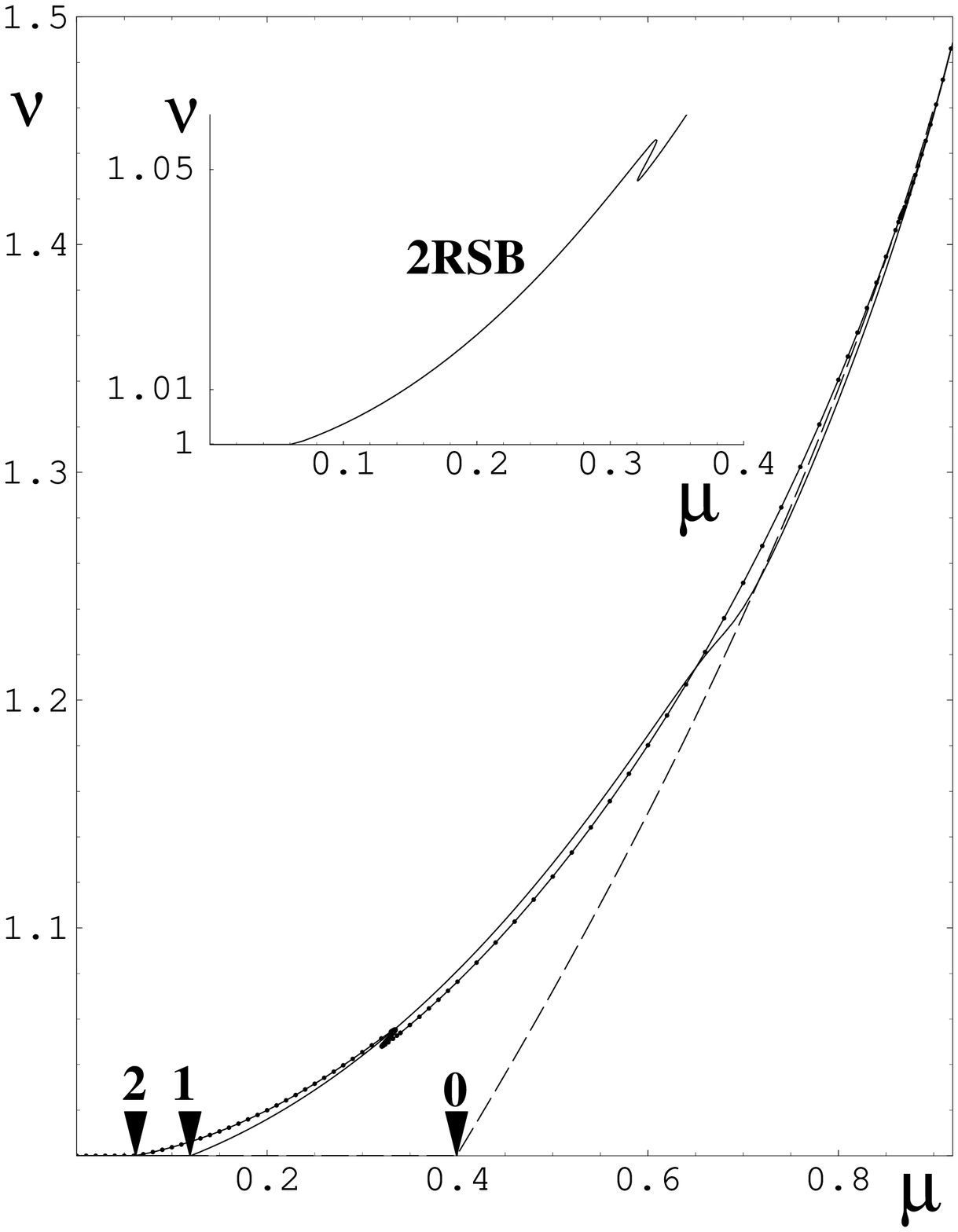}
}
\caption{On the left side the Figure shows the selfconsistent solutions for
$T$-normalized Parisi parameters $a_1, a_2$ and for the order parameters
$q_1,q_2,q_3$ ($\tilde{q}=1$ not shown) at $2RSB$-level as a function of the
chemical potential $\mu$; on the right the fermion filling factor $\nu(\mu)$ is
displayed, where labeled arrow-heads indicate the order of the approximation and the
gap-size related limits of the half-filling regime at
$|\mu|<\bar{\chi}^{(k=0,1,2)}(\mu)/2$.
The small region of a multi-valued filling factor
is zoomed in the inset.
}
\label{figs:a.q.nu}
\end{figure}

\begin{figure}
\centerline{
\epsfxsize9.cm
\epsfbox{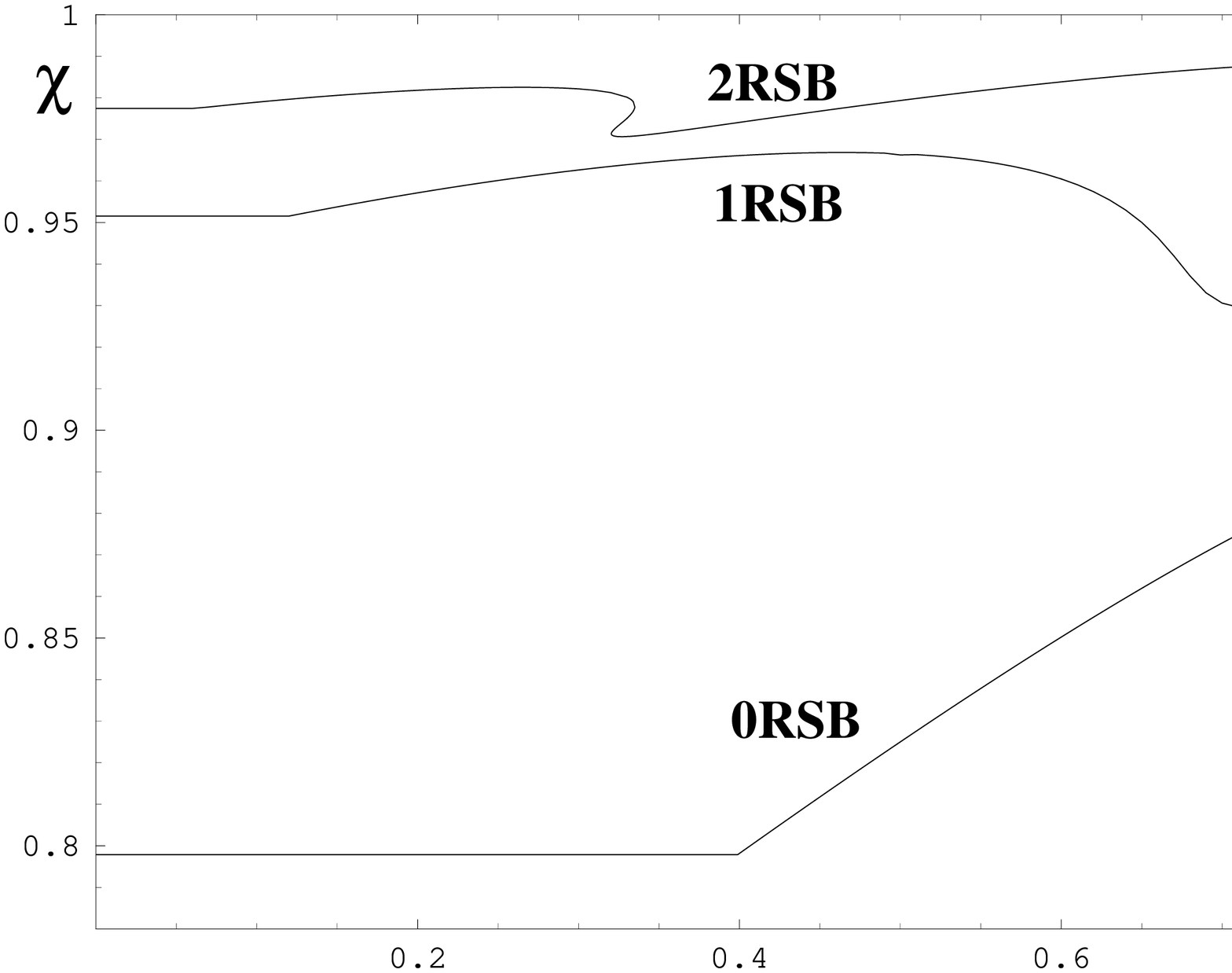}
\epsfxsize9.cm
\epsfbox{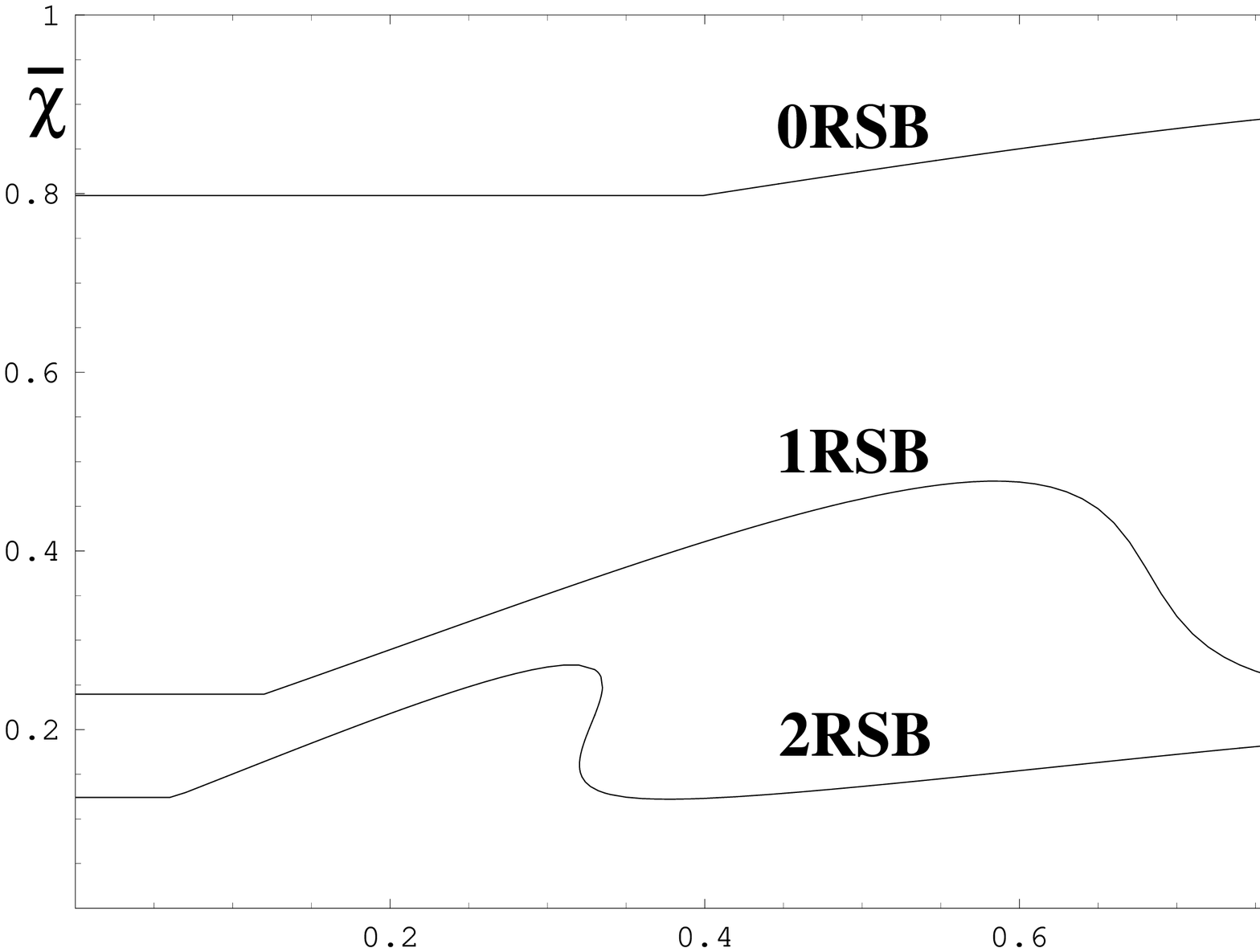}
}
\caption{The equilibrium susceptibility $\chi$ and single-valley susceptibility
$\bar{\chi}$ are shown in 0RSB, 1RSB, and 2RSB approximation at zero temperature.
The constant parts extend over the half-filling regime (at $T=0$) up to $\mu_{_{hf}}$
given by $\bar{\chi}/2$.}
\label{fig:susc.and.susc1}
\end{figure}

The extremization of the free energy at finite temperatures is too complicated to
have yet led to an analytical solution. The $T=0$-limit however allows one
to reduce the number of integrations.
This facilitates a low T-expansion and eases the numerical evaluation.
We take advantage of this simplification and derive the complete set of $T=0$
quantities. The $T\rightarrow 0$-limit leads to the following energy $F(0)$
(per site and in units of $J$)
which, unlike the thermal $F(T)$, depends on three (instead of four) order parameters
$q_{1,2,3}$, two Parisi parameters $a_{2,3}\equiv m_{2,3}/T$, and the so-called
single valley susceptibility $\bar{\chi}$. Extremizing
\begin{equation}
F(T=0)=\frac12 \bar{\chi}(q_1-1)+\frac14 a_1(q_1^2-q_2^2)+\frac14
a_2(q_2^2-q_3^2)-\frac{1}{a_2}\int_{z_3}^G ln \int_{z_2}^G C_0(z_2,z_3)
\label{Eq:F.T0}
\end{equation}
where
\begin{eqnarray}
& &C_0(z_2,z_3)=\frac12 \sum_{\lambda=\pm 1}e^{a_1(\lambda\hspace{.1cm}
H_{eff}(0,0,z_2,z_3)+a_1
(q_1-q_2)/2)}\{1-erf(\frac{
\delta\mu\hspace{.1cm}\Theta(\delta\mu)
-\lambda\hspace{.1cm}
H_{eff}(0,0,z_2,z_3)-a_1(q_1-q_2)}{\sqrt{2(q_1-q_2)}}\}\nonumber\\
& &\hspace{2cm}+\frac12 e^{a_1
\delta\mu\hspace{.1cm}\Theta(\delta\mu)}\sum_{\lambda}\lambda\hspace{.1cm}
erf(\frac{\lambda\hspace{.1cm}
\delta\mu\hspace{.1cm}\Theta(\delta\mu)
-H_{eff}(0,0,z_2,z_3)}{\sqrt{2(q_1-q_2)}}),
\hspace{1cm}\delta\mu\equiv\mu-U-\bar{\chi}/2
\label{Eq:J}
\end{eqnarray}
with $H_{eff}(0,0,z_2,z_3)=\sqrt{q_2-q_3}\hspace{.2cm}z_2+\sqrt{q_3}\hspace{.2cm}z_3$
leads to six coupled (double-)integral equations
\begin{equation}
\{\partial_{a_1},\partial_{a_2},\partial_{q_1},\partial_{q_2},\partial_{q_3},
\partial_{\bar{\chi}}\}F(0)=0
\label{Eq:sceqs-2rsb}
\end{equation}
where $\partial_x$ means $\frac{\partial}{\partial x}$
(Eq.\ref{Eq:F.T0} holds for the $\mu-U/2\geq0$-side particle hole symmetry and up to
irrelevant constants). We determine numerically all parameters for $\delta\mu\geq0$
from the $T=0$ selfconsistency equations which result from Eq.\ref{Eq:sceqs-2rsb}.
The results for $U=0$ and half-filling are shown in the Figures below and all
finite-$U$ results can be obtained from it by the transformation
$\mu\longrightarrow\mu+U$ for positive $\mu$. We shall use the $2RSB$-solutions to
sort out artifacts of the approximation and understand the true underlying physics.\\
Figs.\ref{figs:a.q.nu} and \ref{fig:susc.and.susc1} display new singular features in
$2RSB$ which, by the help of a comparison with the $0$- and $1RSB$-results,
can be suspected to be artifacts of the approximations; they nevertheless provide
an improved insight into the shape of the exact Parisi solution in $\infty RSB$.
The analytic $T=0$-equations explain the origin of $k$ wiggles for $k$-stepped $RSB$.
The surprising appearance of singular wiggles at smaller $\mu\approx\frac13$ involves
multi-valued solutions and, similarly to what happens in first order phase transitions,
would imply discontinuous behaviour. Finally one arrives at the conclusion that
the wiggly behaviour must be associated with - and in fact is caused by -
the stepped approximation of the continuous Parisi function in replica space.
A physical meaning of these solutions could yet remain due to the fact that the
$\infty RSB$ equilibrium solution is approached very slowly in time: finite-step
$RSB$-solutions marked by the single- and two-valley susceptibilities
might be relevant for non-equilibrium behaviour.
%
This discontinuous behaviour will be removed by thermal smearing in $2RSB$ down to
much lower temperatures than observed in Fig.\ref{fig:1rsb-vs} or in the $1RSB$-gap.
The latter was seen to produce a smooth function like the expected exact solution.
Throughout the paper we have chosen the chemical potential to be the independent
variable. In order to see whether the $\mu$-dependence is particularly susceptible
to the soft steps, we inverted $\nu(\mu)$ into $\mu(\nu)$ and obtained the physical
quantities as a function of the fermion concentration $\nu$ as independent variable.
It turns out that the wiggles seen in $\nu(\mu)$ itself do not remove (compensate)
these features from the order parameters and susceptibilities
$q_i(\nu),\chi(\nu),\bar{\chi}(\nu)$.
\begin{figure}
\epsfxsize=14cm
\centerline{
\epsfbox{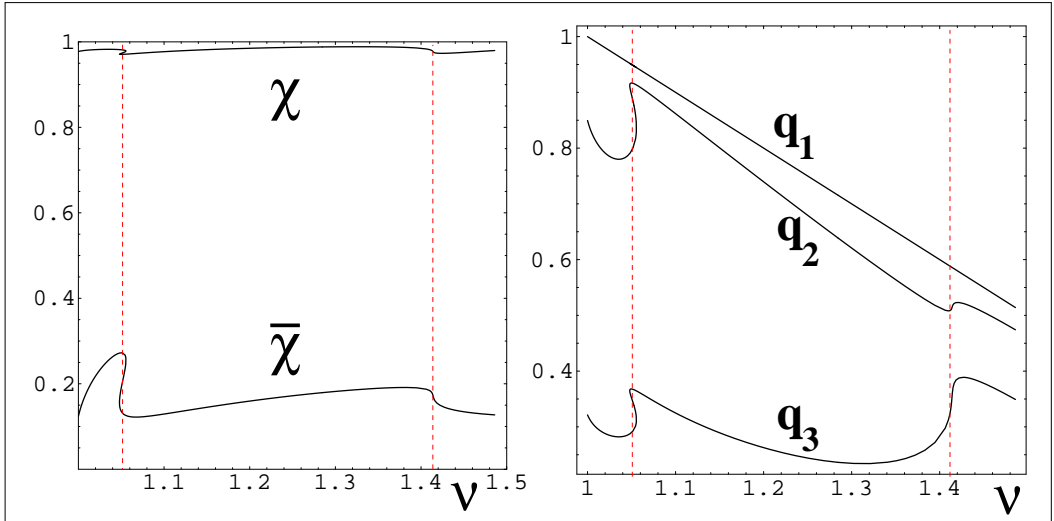}
}
\caption{Order parameters and susceptibilities in as a function of
fermion concentration $\nu$ in $2RSB$-approximation. Vertical dotted lines locate
wiggly behaviour generated by the Parisi steps in replica space.}
\label{q-of-nu}
\end{figure}
Although some unphysical features are still contained in the $T=0$-solutions at
any finite $RSB$, $2RSB$-solutions are much closer to the exact ones.

Since the Parisi function $q(x)$ is constructed by piecewise continuous/constant
functions of $x$ ($\theta$-functions below a break-point), the present results
indicate that this dependence in replica space transforms either into strong variation
or into jumps which resemble first order phase transitions. Their positions on
the $\mu$-axis scale with the step positions along the replica axis (replica
dimension). A real discontinuous phase transition occurs however at the magnetic
breakdown value $\mu\approx 0.88$ for $U=0$ and shifted by $U$ as shown in
Figure \ref{Eq:Tf}. The critical filling is independent of the Hubbard coupling and
roughly given by $1.45$.\\
By inspection of the selfconsistent equations one expects that the number of wiggles
will increase with the number $k$ of RSB-steps; the size of the
jumps will diminish, thus improve the quality of the approximation, and finally
disappear in the $k\rightarrow\infty$-limit.
\subsection{The order parameter as a function of chemical
potential and $T=0$-measure of replica space}
The $2RSB$-results for the discretized order parameter and the intervals $a_k$ of
their step heights given in Figure \ref{figs:a.q.nu} can be combined and represented
in one Figure. An instructive representation of the generalized
Parisi-function then results in the space of $\mu$ and of a variable called $a_x$.
The latter replaces the variable $x$ used by Parisi to represent the solution
$q(x)$. In Parisi's papers, the continuous parameter $x$ was
introduced to host all matrix block size parameters $m_k$. These
are found within an interval $0\leq x\leq1$ in the replica limit (in
addition they lie dense in the $\infty RSB$ limit). The function $q(x)$ is
discretized by means of $\theta$-function (steps) in any finite order of $RSB$.
In case of the low temperature limit, the selfconsistent solutions for matrix block
size parameter $m_k$ turned out to vanish linearly with temperature, which
required the use of a rescaled parameter $a_k\equiv m_k/T$. These
parameter values are now hosted in an unbounded interval and we
consider them as discrete values on an $a_x$-axis, where $a_x$
replaces the low temperature limit of $x/T$ in a $T=0$
calculation.
Without temperature-rescaling the $x$-interval below the break point, where the
nontrivial part of the Parisi function shows up, would collapses to a point with
q(0) infinitely-valued. The nontrivial part of the Parisi function still exists at
$T=0$ and is well represented on the $a_x$-axis. As a function of $a_x$ and of
the chemical potential we obtain the $2RSB$-result of
$q(a_x,\mu,T=0)$ as given by Fig.\ref{q2rsb_3d}\\

Figures 10 and 11 appended (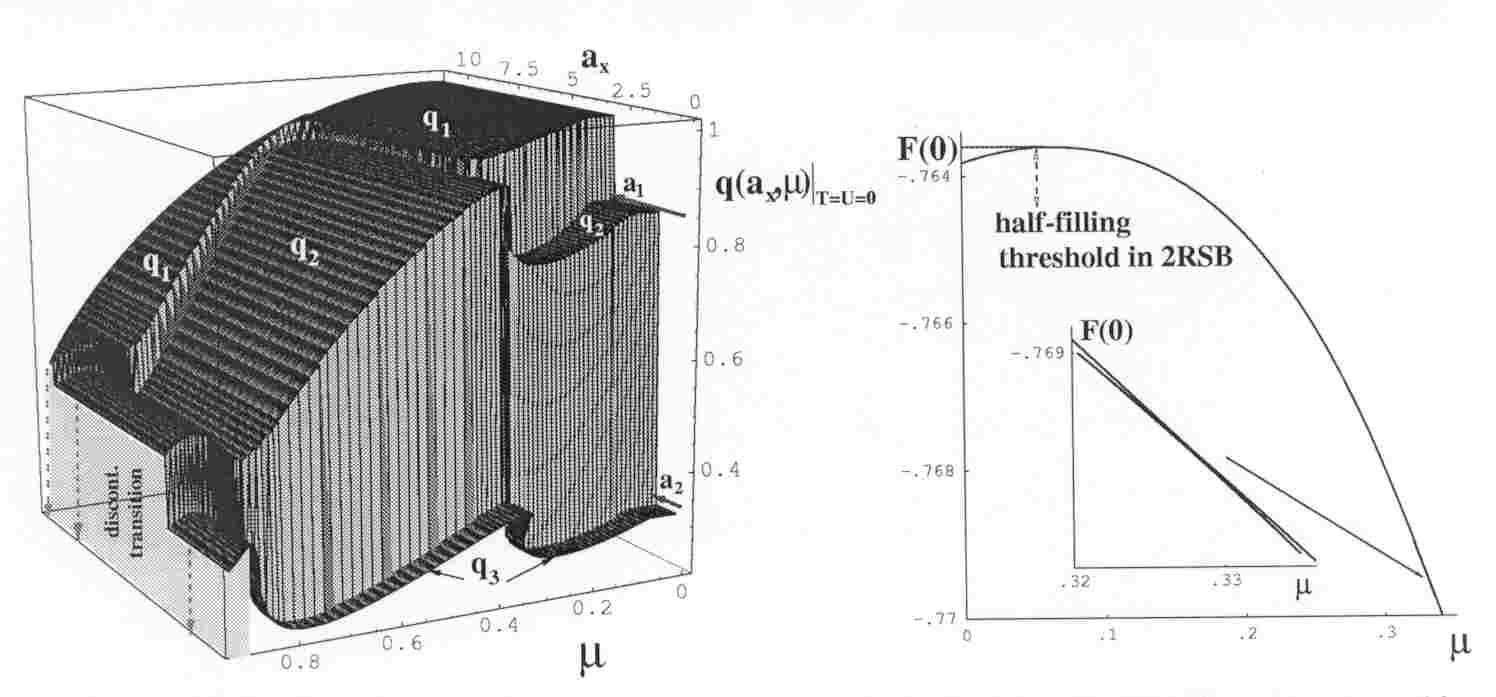) at e.o.f.
\begin{figure}
\caption{$2RSB$-approximation of the $T=0$ Parisi order parameter as a function
of $a_x\equiv x/T|_{T=0}$ in analogy with
$a\equiv lim_{_{T\rightarrow0}}m(T)/T$ and $\mu$.
The strongly $\mu$-dependent interval lengths over which the order parameter assumes
one of its possible three constant values are shown at the $\mu=0$-side of the
plot. At the large $\mu\approx0.88$-side the first order regime of magnetic breakdown
is shown.}
\label{q2rsb_3d}
\caption{The calculated $T=0$ limit of the free energy in $2RSB$. For
$0\leq\mu\le\bar{\chi}/2$ the half-filling energy is higher
(dashed line). The inset show the multi-valued regime indicated by
an arrow in the large-scale energy.}
\label{fig:2rsbenergy}
\end{figure}
At $T=0$ it seems possible that the solutions as a function of $\mu$
for very high $k$ (close to the exact one at $k=\infty$)
will be made up of piecewise continuous functions of $\mu$, where
discontinuities at the $O(k)$ wiggles and the intervals of continuous behaviour
both approach zero. Some resemblance with level crossing/repulsion appears.
The wiggles must finally be seen to mark a soft stepping of the order parameter
in contrast to (and as a consequence of) the hard steps along $a$ which are
imposed by the definition of the Parisi scheme.
One may be tempted to ask for a generalized scheme that softens the steps also
in the replica dimension before the exact limit is reached. This seems even more
desirable since the higher order $RSB$-calculations below show an extreme
flatness of the energy landscape.\\
Using the above solutions we calculate the $T=0$-limit of the free energy of
Eq.(\ref{Eq:F.T0}). The solution shown in Fig.\ref{fig:2rsbenergy} confirms the
half-filling limit at $\mu=\bar{\chi}/2$ and a discontinuous transition
near $\mu\approx 0.33$.
Thus it is clear that the step-approximation of the order
parameter function in replica space leads either to strong
variation or to jumps, which appear in the form of artificial first order
transitions at certain values of the chemical potential; these locations
correspond to and scale with the positions of the Parisi-steps.
No such transition can be expected in the exact solution.
One could speculate whether $\mu=0$ is the accumulation point of these discontinuities
and thus may be the source of special critical behaviour in the $\mu=0$ point in
$\infty RSB$. In this case infinitely many wiggles could render the Parisi function
a mathematically delicate object at half-filling. While Parisi's
finite temperature order parameter function $q(x)$ is defined on
an interval $0\leq x\leq1$ the present replica interval for $T=0$
is unbounded and there is not yet an indication that the discontinuities
must lie dense in the limit of infinitely many $RSB$-steps.\\
The conclusion of this chapter is that the $\mu$-dependence seen
in the susceptibilities will disappear and yield
$\chi=1,\bar{\chi}=0$, while the order parameter function
decreases smoothly as $\mu$ approaches the magnetic breakdown
value and steps along the $a_x$-axis must be continuized. In this
sense Figure \ref{q2rsb_3d} offers a crude picture which calls
for an improved and refined treatment. Before this is achieved in
section \ref{section:4RSB} the fermion filling $\nu(\mu)$, which is related
for $T=0$ to the largest order parameter $q_1$ and also to the integrated
density of states, is obtained.
\subsection{The $\nu(\mu)$-dependence from 2$RSB$-calculation for the whole
spin glass regime}
\begin{figure}
\centerline{
\epsfxsize=12.cm
\epsfbox{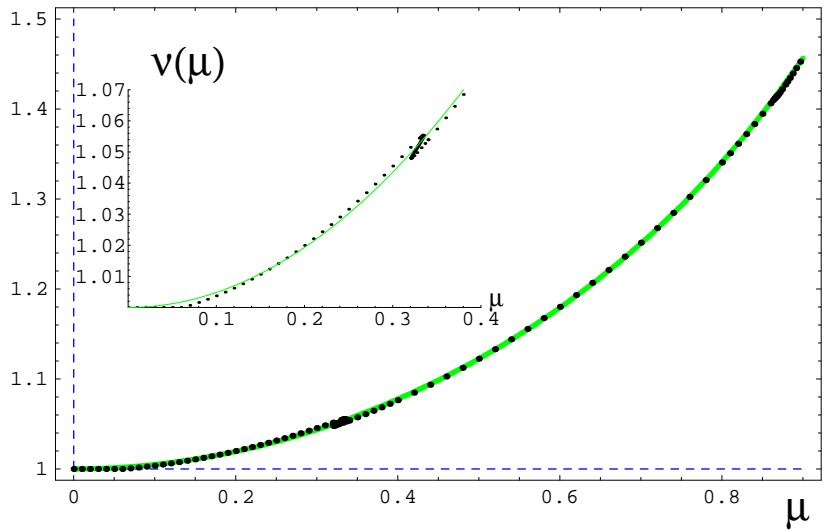}
}
\caption{The known exact value at half-filling is joined with the
$2RSB$-approximate values of the fermion filling factor as a function of the
chemical potential. The $RSB$-artificial wiggles are almost innocuous on this scale.}
\label{nu-fit}
\end{figure}

%
The spin glass regime exists at $T=0$ in the range
$0\leq|\mu-U/2|<\mu_c\approx0.9+U/2$ for positive $U$.
Comparing the $kRSB$-results for $k=0,1,2,3,4$ together with the known exact value at
$\mu=0,T=0$ the regime of validity and the quality of $2RSB$ can be estimated.
An interpolation helps to conjecture the form of the expected exact
solution$\nu(\mu)$.
The numerical result can also be used to generate an analytical fit, which in the best
case would be the exact solution. If not exact, it can still represent an analytical
form for the propagator needed in diagram expansions of itinerant models for example.
Figure \ref{nu-fit} shows such an interpolation.
The fermion concentration $\nu=\int^{\mu}_{-\infty} d\epsilon\rho(\epsilon)$
at $T=0$ equals $2-q_1=2-q^{aa}$.
Only the zoomed inset in Figure \ref{nu-fit} shows significant deviations
between the $\nu(\mu)$ fit-function and the $2RSB$-calculated points.
Viewed more globally over the whole $\mu$-interval a simple quadratic behaviour
with small quartic corrections fits nicely.
The exact behaviour of $\nu(\mu)$ appears to be simple.
The fit function, which interpolates between $\nu_{exact}(\mu=0)=1$ and
$2RSB$-values for $\mu>0.12$, is close to $1+\frac12 \mu^2+O(\mu^4)$.
The way in which the finite half filling limits ($\nu=1$) move to
smaller chemical potentials is determined in the following section
\ref{section:4RSB}, confirming the interpretation $\nu\sim\mu^2$.
\section{The shape of the spin glass pseudo-gap, its
combination with the Hubbard gap, and a comparison with the
Efros-Shklovskii Coulomb pseudogap:\\
4RSB-calculation at half filling, exact relations}
\label{section:4RSB}
We extend the above-mentioned extremization of the free energy,
Eq.\ref{Eq:sceqs-2rsb}, to a $4RSB$-solution.
On this level and for arbitrary filling, the set of ten coupled
integral equations, each having seven coupled integrations,
\begin{equation}
\{\partial_{a_1},\partial_{a_2},\partial_{a_3},\partial_{a_4},
\partial_{q_1},\partial_{q_2},\partial_{q_3},\partial_{q_4},\partial_{q_5},
\partial_{\bar{\chi}}\}F(0)=0
\label{Eq:sceqs-4rsb}
\end{equation}
must be solved selfconsistently. We performed this calculation for half filling
where the numerical efforts are reduced to the solution of eight
coupled equations, since $q_1=1$ and the $\bar{\chi}$-equation
decouples from the rest. If the points in replica space, where the
order parameter values are sought, would not be required to
minimize the free energy too one could include several higher $RSB$-orders
with similar effort. This seems however unnecessary for the present purpose and
hence was discarded.
\subsection{4-step RSB approximation for $T=0$-order parameter
function and -susceptibilities at half filling}
While a Parisi order parameter function for finite temperatures is
defined on an interval $0<x<1$, the $T=0$ order parameter function
is represented by the temperature rescaled quantity
$a_x\equiv\lim_{T\rightarrow0}x/T$, which is the natural choice
due to the $m(T)\sim T$ dependence observed in our selfconsistent calculations.
An upper bound for the selfconsistent $a_x$-solutions does not seem to exist.
The maximum value in $4RSB$ is $5.5$ and grows rapidly with the order of $RSB$.
A mapping $a_x\rightarrow\tilde{a}_x\equiv a_x/(1+a_x)$ has the virtue of
re-converting even the $T=0$ interval to $0\leq\tilde{a}_x\leq1$.
The following Figures show the obtained selfconsistent values and
the fitting continuous order parameter function at $T=0$. The
fitting analytical form indicates that a distribution of error
functions is probably involved in the exact analytical solution of
the problem.
\begin{figure}
\centerline{
\epsfxsize11cm
\epsfbox{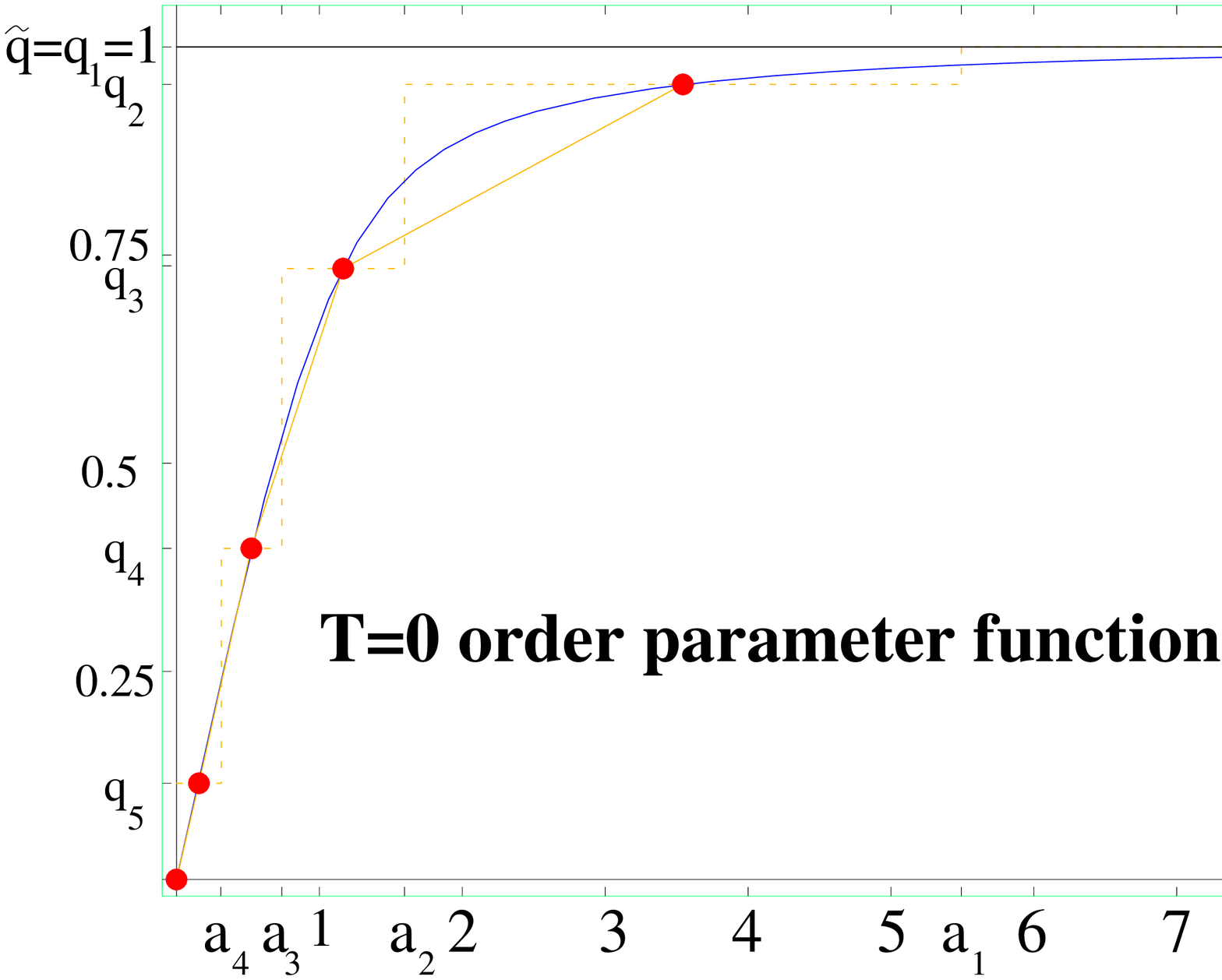}}
\centerline{
\epsfxsize11cm
\epsfbox{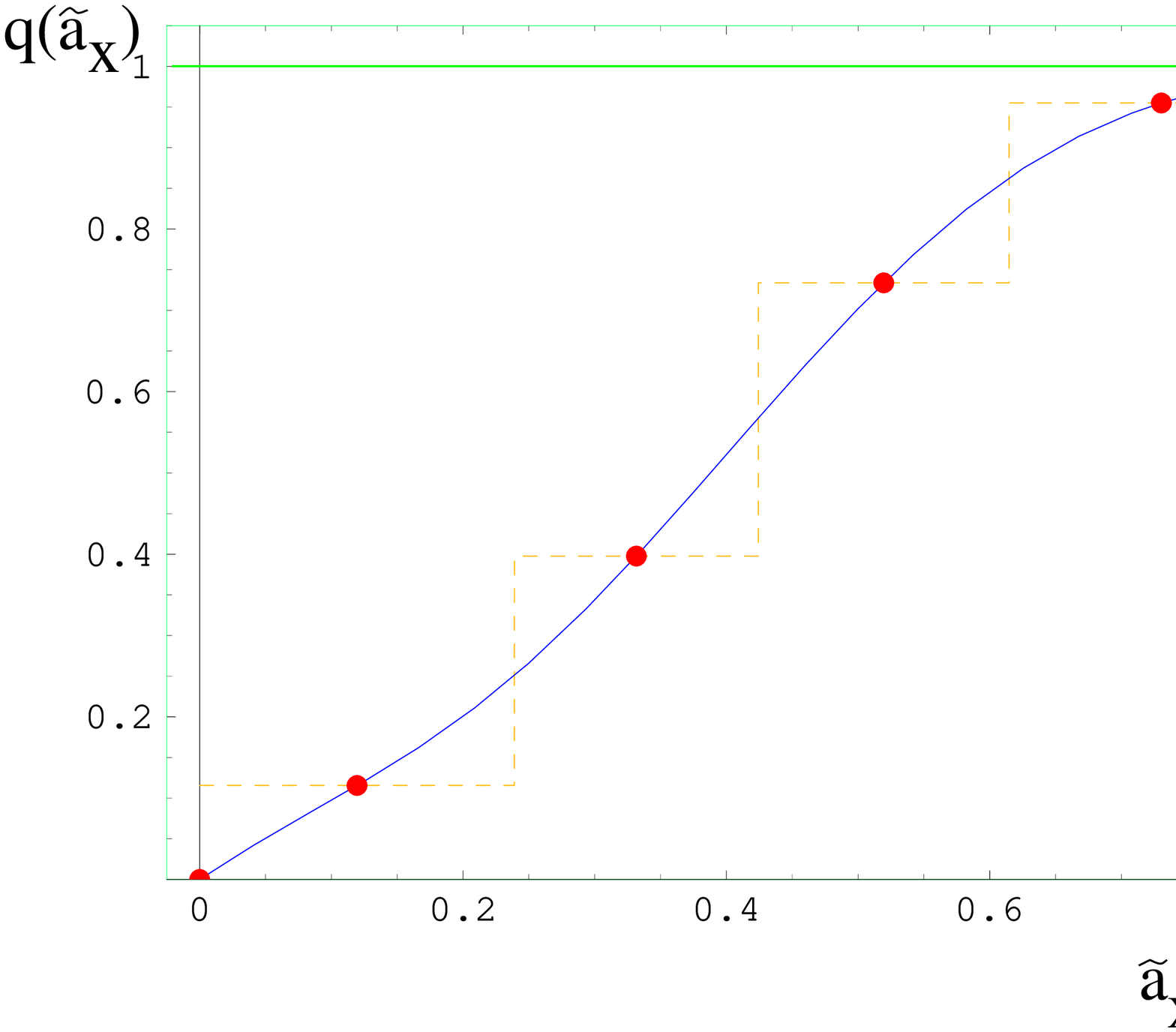}
}
\caption{A $T=0$ order parameter (analytic model) function, which fits well the
calculated 5-step ($4RSB$) approximation, is shown. The upper Figure displays
$\{\tilde{q}=q_1=1,q_2,...,q_5\}$ and $\{a_1,...,a_4\}$, all
obtained selfconsistently, for the original variable $a_x$.
Below, the same function is shown as a function of $\tilde{a}_x$ after mapping to
the interval $[0..1]$ by means of $a_x\rightarrow a_x/(1+a_x)$.
}
\label{Fig:4rsb-order-parameter}
\end{figure}
A break point, which is a specific in finite temperature Parisi
order parameter function $q(x)$, appears to be absent in
accordance with the expectation: the break point scales like
$x_{b.p.}\sim\sqrt{T}$ and since $a_x\sim x/T$ one expects this
point to move to infinity in the zero temperature limit.\\
The equilibrium and nonequilibrium susceptibilities, obtained from
our $4RSB$-calculation are shown in
Figure\ref{Fig:4rsb-susceptibilities}.
The $\bar{\chi}$-decay is well modeled by a function which decays
like $1/k$ when the number of $k$ $RSB$-steps goes to infinity.
Logarithmic corrections or a slower decay cannot yet be ruled out.
\begin{figure}
\centerline{
{\epsfxsize8.cm
\epsfbox{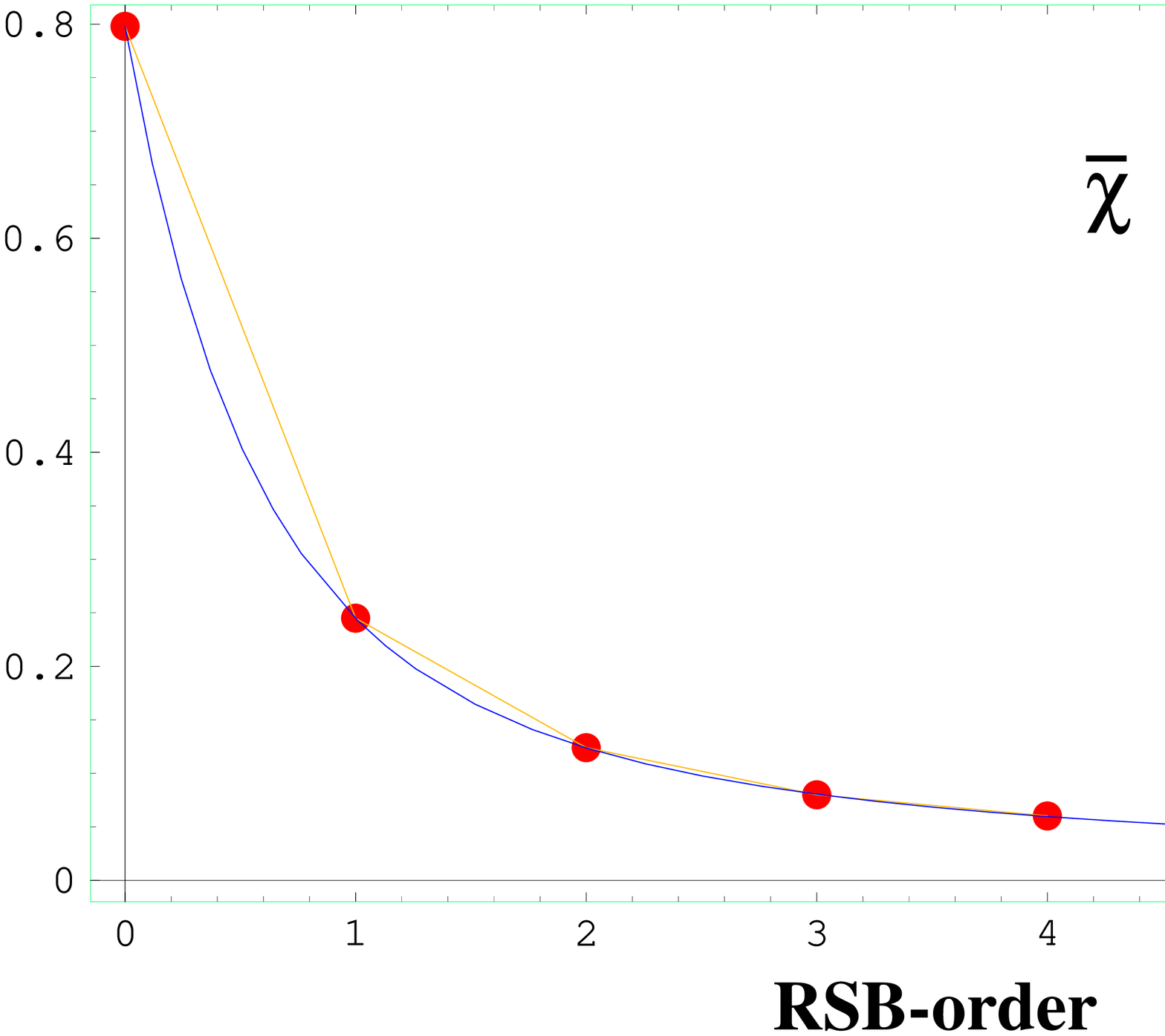},
\epsfxsize8.cm
\epsfbox{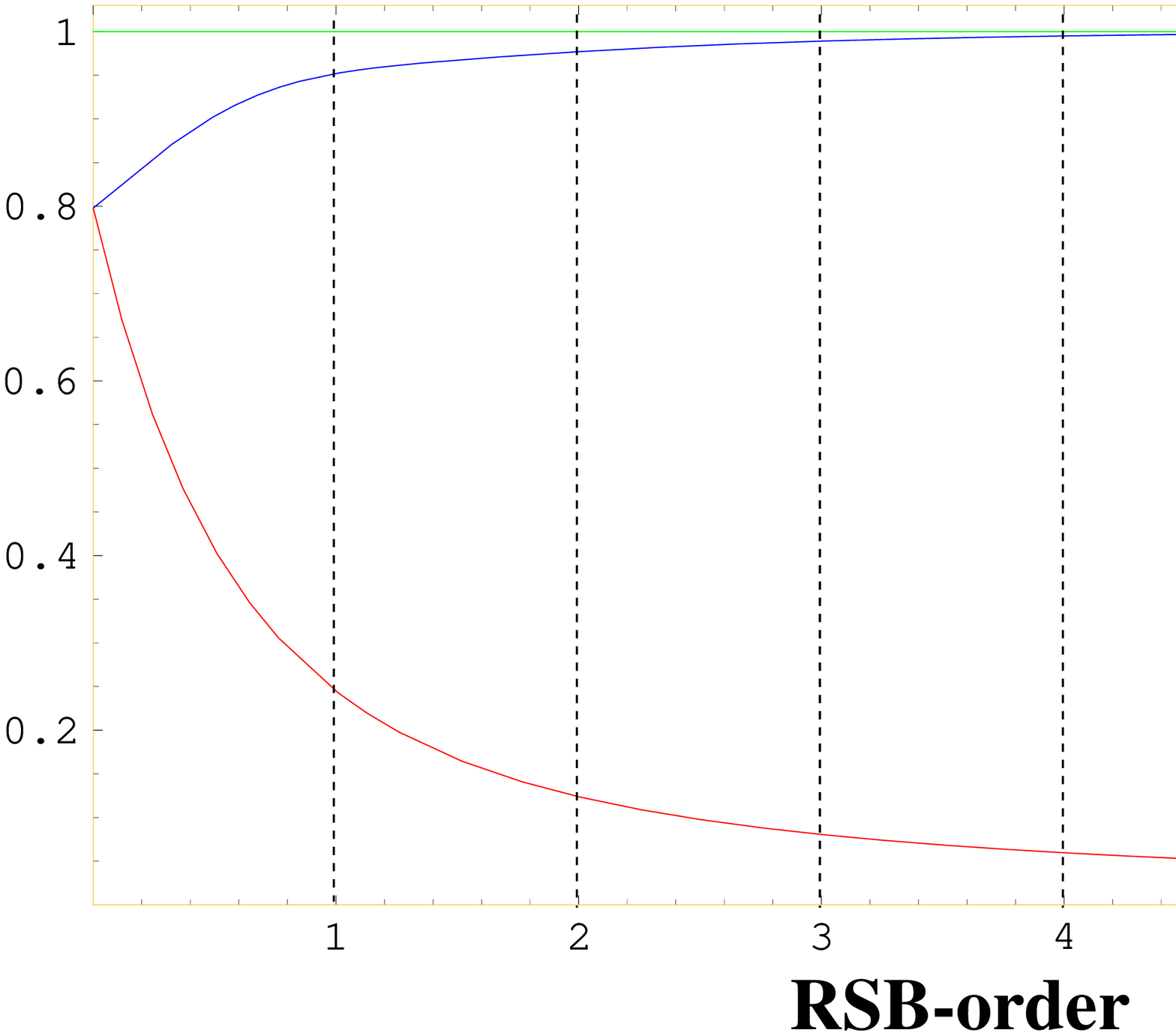}}
}
\caption{The decay of the single-valley (nonequilibrium) susceptibility
$\bar{\chi}$ is shown as a function of interpolated $RSB$-order; calculated
points are shown by dots and extrapolation to higher order is possible from
a simple and good analytical model fit. The growth of the equilibrium susceptibility
$\chi$ towards $1$ is shown in comparison with the $\bar{\chi}$-fit in the Figure
below.}
\label{Fig:4rsb-susceptibilities}
\end{figure}
\subsection{4$RSB$-approximation of the spin glass pseudogap}
\begin{figure}
\centerline{
\epsfxsize14.5cm
\epsfbox{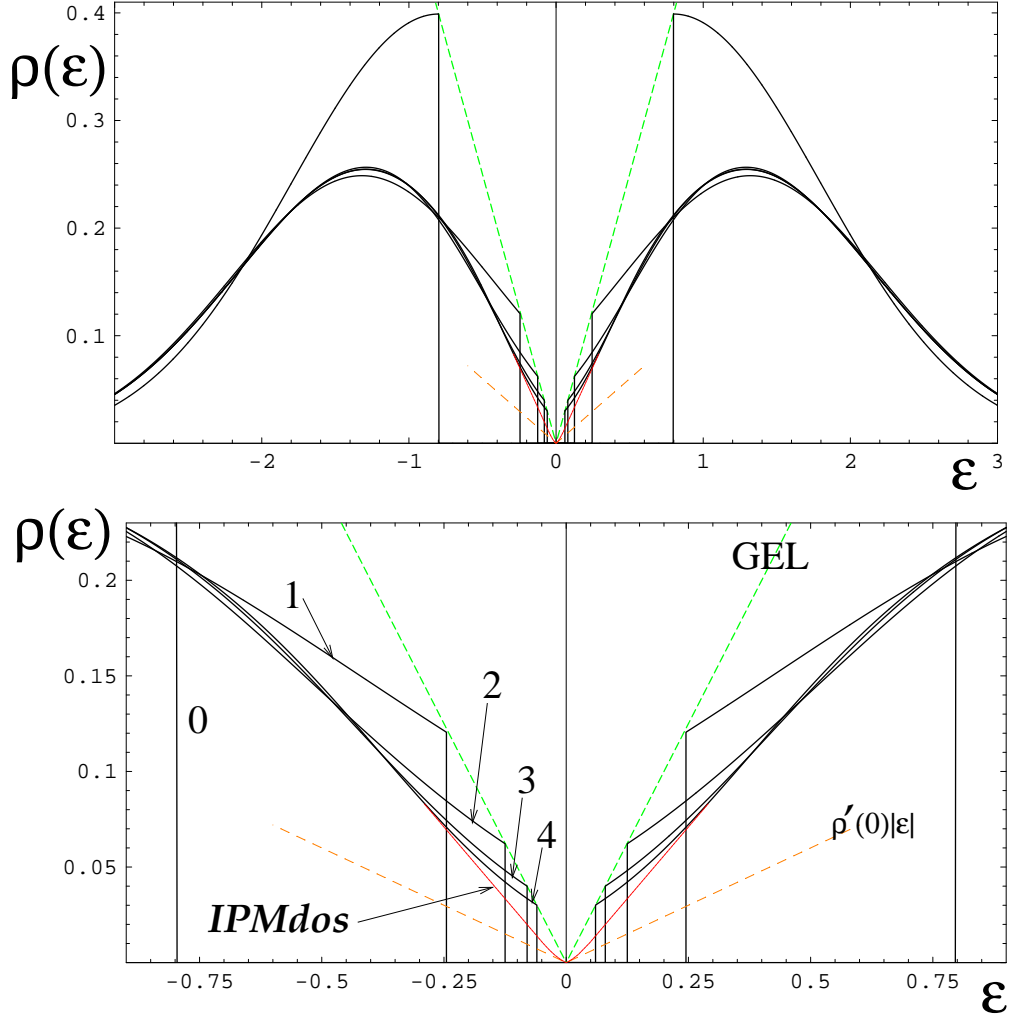}
}
\caption{The magnetic band structure at $T=0$ and for half-filling
is shown for comparison in the replica-symmetric (0RSB), 1-, 2-, 3-, and 4-step
replica symmetry breaking approximation (solid curves labeled with
$0,1,2,3,4$ correspondingly). The gap energies decrease with the
increasing $RSB$-order and the shape of the gap emerges. The wide scale upper Figure
shows the decrease of width and position of energy regimes, where the next higher
order of $RSB$ bifurcates.\\
The bottom Figure zooms the gap regime near $E_F$. The dashed gap-edge-line 'GEL'
shows the exact lines connecting $\rho(E_g)$ ($E_g$ denoting the gap edge energies)
while the asymptotic power law in the limit $\epsilon=0$ is indicated by the
dashed line $\rho'(\epsilon=0)|\epsilon|$ with $\rho'(0)$
approximated by its $4RSB$-value $0.13$. An interpolation $IPMdos$ of the
density of states between the tiny gap-bottom regime following this law and
the wide gap regime (where $4RSB$ is almost exact) with a slope closer to
$0.3$ is modeled by extrapolating a $\rho''$-calculation to $\infty RSB$.
}
\label{Fig:4rsb-hf-dos}
\end{figure}

Thermal effects cover unphysical remainders of replica symmetry and, as one
increases the order of $RSB$, the reappearance of incorrect features shifts to ever
smaller temperatures. In a comparable way one needs higher orders of $RSB$
to describe exactly smaller and smaller excitation energies above
(and below) the Fermi level. In other words, $\epsilon\equiv|E-E_F|$ scales with $T$
and the larger the excitation energy is the smaller is the effect
of the 'fine structure' of the Parisi function for small $a_x$ as
shown in Figure \ref{Fig:4rsb-order-parameter}.
The effect can be seen in the fermionic density of states $\rho$ at $T=0$.
It will also enter the quantum-dynamical fermion propagator, which is determined by
its spectral weight $\rho$. We improved earlier one-step $RSB$-calculations
\cite{rohf} by three orders and find that four-step $RSB$ ($4RSB$) is sufficient
to predict the shape of the pseudogap.\\
We derive the exact $4RSB$-formula for the fermionic density of states
$\rho(E\equiv\epsilon+\mu)$ for $T=0$ as
\begin{eqnarray}
& &\rho(E)=e^{a_1(E-\bar{\chi})}\frac{1}{\sqrt{2\pi(q_1-q_2)}}
\int_5{\cal{N}}_2(a_4/a_3)
\int_4{\cal{N}}_2(1-a_4/a_3)
\int_3{\cal{N}}_1(1-a_3/a_2)
\int_2
e^{-(E-\bar{\chi}-H_{eff})^2/(2(q_1-q_2))}C^{a_2/a_1-1}
\nonumber\\
\label{Eq:4rsbdos}
\end{eqnarray}
using the integral operator
$\int_b\equiv\int_{z_b}^G\equiv\int_{-\infty}^{\infty}dz_b
\hspace{.1cm}exp(-z_b^2/2)/\sqrt{2\pi}$ acting on the accumulated normalizing factors
\begin{equation}
\{{\cal{N}}_1(x)\equiv\left[\int_{2'}
C^{a_2/a_1}\right]^{-x}\quad,\quad
{\cal{N}}_2(x)\equiv\left[\int_{3'} \left[\int_{2'}
C^{a_2/a_1}\right]^{a_3/a_2}\right]^{-x}\}
\end{equation}
and $C$ as given by Eq.\ref{Eq:J} after replacement of the $2RSB$ by the
$4RSB$-effective field $H_{eff}$.
Our numerical evaluation of Eq.\ref{Eq:4rsbdos} for half-filling
and $U=0$ is shown in Fig.\ref{Fig:4rsb-hf-dos}.
The dash-dotted straight lines locate the shoulder-height $\rho(E_{g,k})$
at the gap edges $E_{g,k}$ of $k$-step $RSB$:
the ratio between these heights and the gap-width is invariant ($=1$)
under a change of the order $k$ of $RSB$, ie $RSB$-invariant.
\begin{figure}
\centerline{
\epsfxsize15cm
\epsfbox{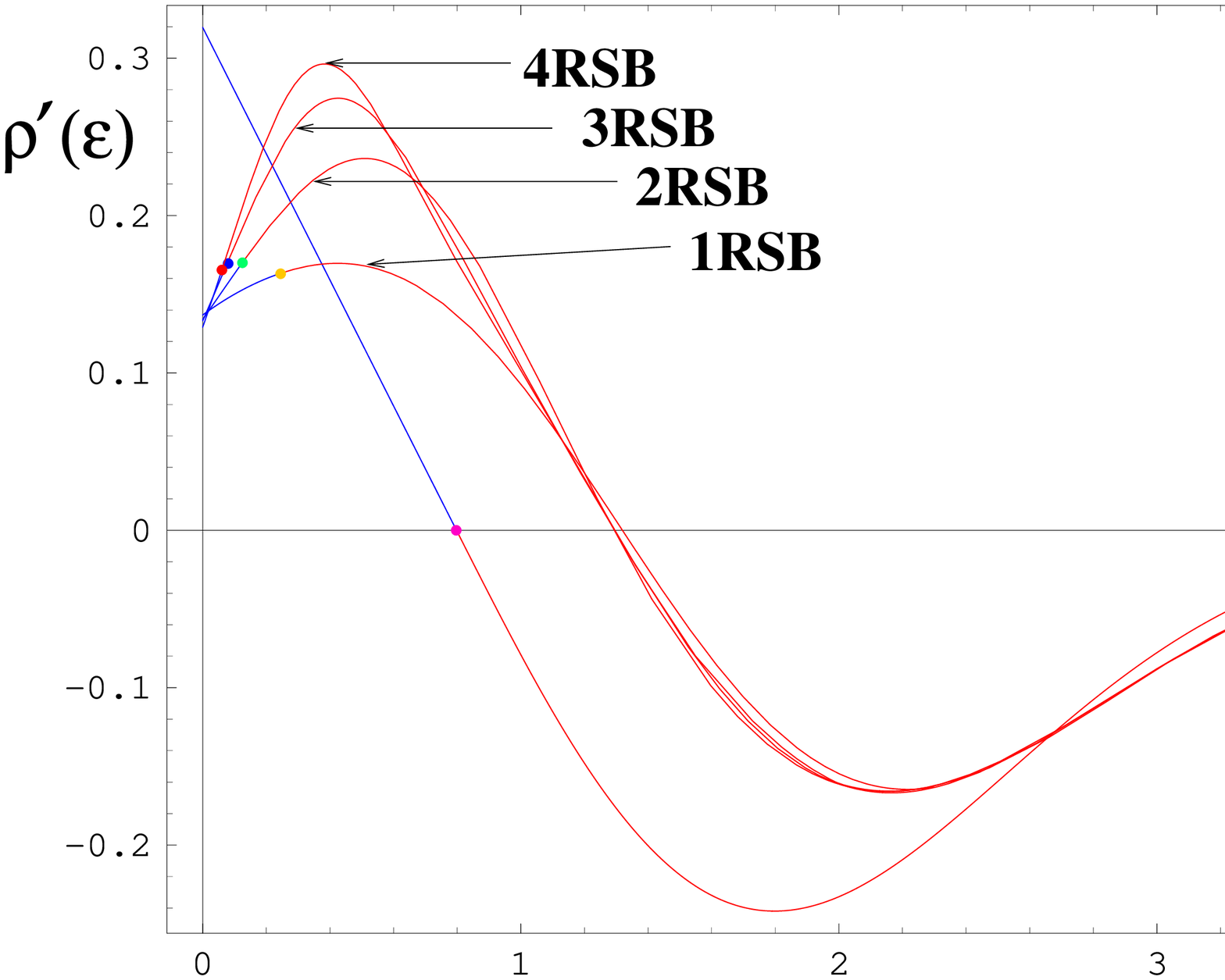}
}
\caption{First derivative of the density of states $\rho(\epsilon)$, $\epsilon\equiv
E-E_F$ from replica-symmetric to four-step $RSB$ approximation. Gap edges are indicated
by points for each approximation and an extrapolation beyond these points towards the
Fermi level ($\epsilon=0$) is shown. These lines demonstrate the rapid convergence:
lines from $2-,3-,$ and $4RSB$ lines meet in the point $(0,0.13+O(10^{-3}))$.}
\label{Fig:d1dos}
\centerline{
\epsfxsize15cm
\epsfbox{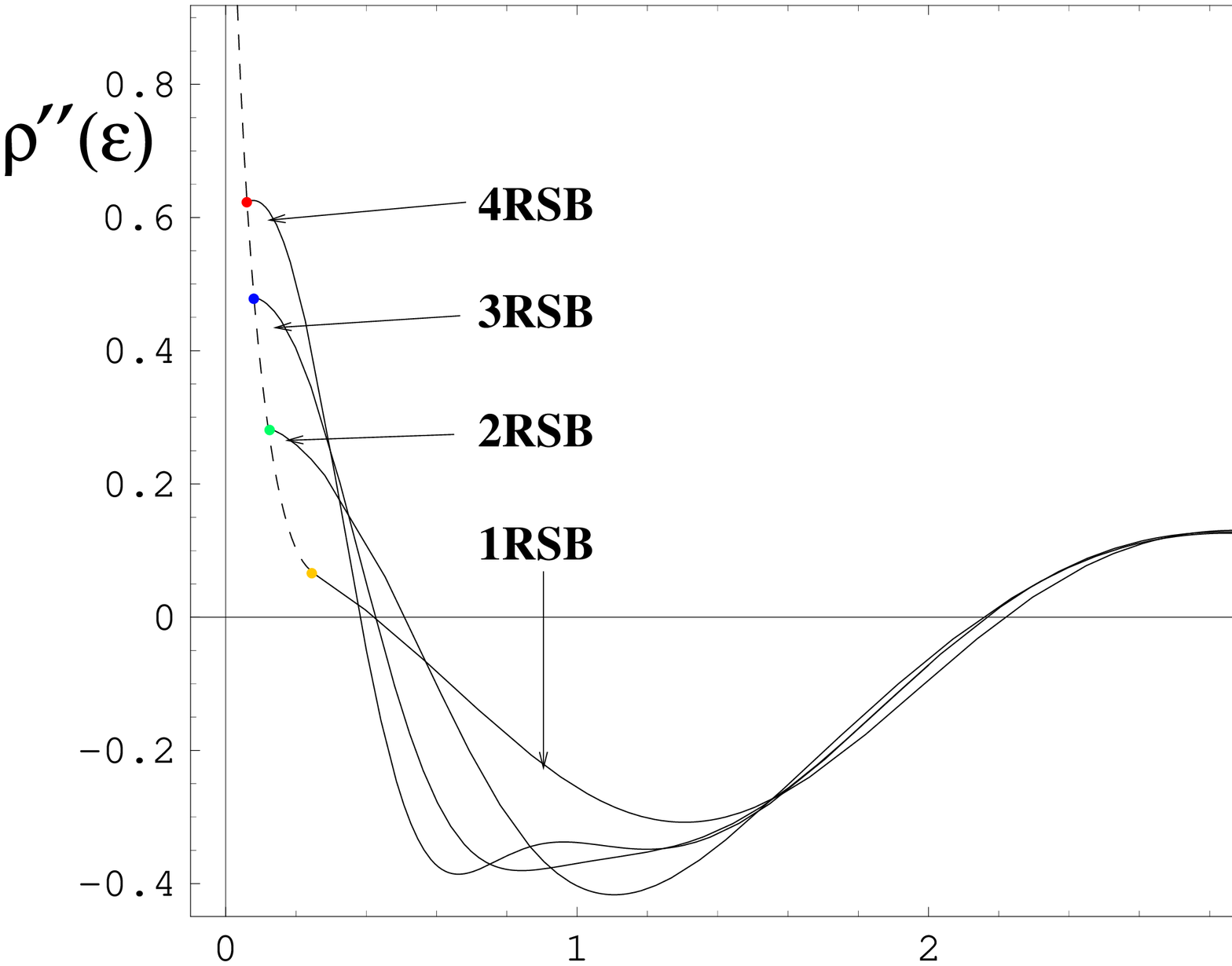}
}
\caption{The second derivatives of the density of states does not show rapid
convergence up to fourth order replica symmetry breaking.
Dots mark again the gap edge positions $\epsilon=E_{g,k}$ in k-th order $RSB$.
Their position is well fitted to the shown curve (dashed)
$\rho''(E_{g,k})=2.1-7.9\sqrt{E_{g,k}}+7.6 E_{g,k}$. This serves as a
crude estimation of $\rho''(0)$ in $\infty RSB$.
A divergence of $\rho''$ can however not be excluded, which would imply the
subleading exponent of $\rho(\epsilon)$ to be smaller than $2$.
}
\label{Fig:d2dos}
\end{figure}
In Figure \ref{Fig:d1dos} the gap edge points $E_{gk}$ must reach $\epsilon=0$
in the $\infty RSB$ limit; whether these values $E_{gk}$ for further increasing
values of $k$ eventually fall on a curve with divergent slope for
$\epsilon\rightarrow0$ can not yet be concluded from Figure \ref{Fig:d2dos}.
This affects only the subleading behaviour which determines the curvature within
the very small asymptotic scaling regime.
Figure \ref{Fig:4-and-infty-rsb-dos-model} illustrates the remaining small
corrections beyond $4RSB$. The interpolating approximation to
the exact solution shows the smallness of the scaling regime,
while the larger linear regime up to $\epsilon\approx1$ can be
viewed as subcritical.
Without a closer look it is hardly possible to see that the extrapolation of the
almost linear DOS-decay would not end in $\epsilon=0$ but leave a tiny finite
fake-gap. This is prevented by the crossover to the scaling regime as described
in the zoomed part of Figure \ref{Fig:4rsb-hf-dos}.\\

Figure 18 (fig18.gif) appended at e.o.f.
\begin{figure}
\caption{The magnetic band structure at $T=0$, for $U=0$, and half-filling.
The $4RSB$-approximation, which contains only a tiny gap, is shown in comparison
with the interpolated model ($IPMdos$) of full $RSB$ matching $\rho(E_F)=0$ and the
$\rho'(E_F)|\epsilon|$-law, concluded from RSB-convergence towards
$\rho'(E_F)\approx0.13$ of Figure \ref{Fig:d1dos}, with the
regime $|\epsilon|>0.3$ where the $4RSB$-values are effectively exact.
}
\label{Fig:4-and-infty-rsb-dos-model}
\end{figure}

{\it Effects of the Hubbard interaction:}
At zero temperature and  for half-filling the effect of the Hubbard coupling
is to spread the spin glass gap by the $U$ and to shift the symmetry point
$\mu\rightarrow\mu-U/2$. The repulsive Hubbard interaction preserves half-filling
and maintains the gap even in the limit of $\infty RSB$.\\

Figure 19 (fig19.gif) appended at e.o.f.
\begin{figure}
\caption{The density of states for vanishing Hubbard interaction
and asymptotic $|\epsilon|$-behaviour is shown in comparison
with one for finite Hubbard repulsion $U/J=1$. At $T=0$ and for half filling
the Hubbard gap splits the spin glass pseudogap in two halfs and
preserves the form of the linear $\rho$-decay relative to the gap.
}
\label{Fig:infrsb-model-U}
\end{figure}
It is perhaps unexpected to evoke universal critical behaviour in the context
of a pseudogap in a range-free (or infinite-dimensional) model. However the results
indicate that critical correlation equivalents appear to exist in replica space.
It is in this space that a theory perhaps of renormalization group character
should be developed in order to analyze the critical behaviour (shape)
of the pseudogap.
In the better known case of finite range interactions (in finite dimensions)
the renormalization group approach as defined and applied to the Coulomb pseudo-gap
by Johnson and Khmelnitskii \cite{khmelnitskii} is a related very
interesting but low-dimensional example.
Renormalization group studies of random magnets of Ref.\onlinecite{dotsenko}
and techniques described in the review by Shankar \cite{shankar}
must be considered to eventually understand the spin glass gap in
presence of long-range Coulomb interaction too.
\subsection{Fermion propagator and spectral representation}
\label{fermion-propagator}
Perturbation expansions, which use the present model as a free (solvable) limit,
are an interesting possibility to study for example itinerant spin glass models.
In such expansions the fermion propagator, which is a site-localized propagator
in time, should be known analytically. In particular its analytical properties
are required in order to be able to evaluate diagrams of the perturbation theory.
While it is not yet clear whether an exact analytical solution can be found in the
$\infty RSB$-limit, the present analysis allows to approach this solution in a
qualitative way. The numerical study helps to find an analytical fit function,
which could be considered as a diagrammatic element in the expansions mentioned above.
Using our previous result for the density of states $\rho(\epsilon)=
-\frac1\pi Im(G^R(\epsilon))$ the spectral representation
\begin{equation}
G(i z_n)=\int_{-\infty}^{\infty} du\rho(u)/(iz_n-u)
\label{spectralrep}
\end{equation}
with $z_n=(2n+1)\pi k_B T/\hbar$
allows to evaluate the real part and thus the full Green function too.
\begin{figure}
\centerline{
\epsfxsize=15cm
\epsfbox{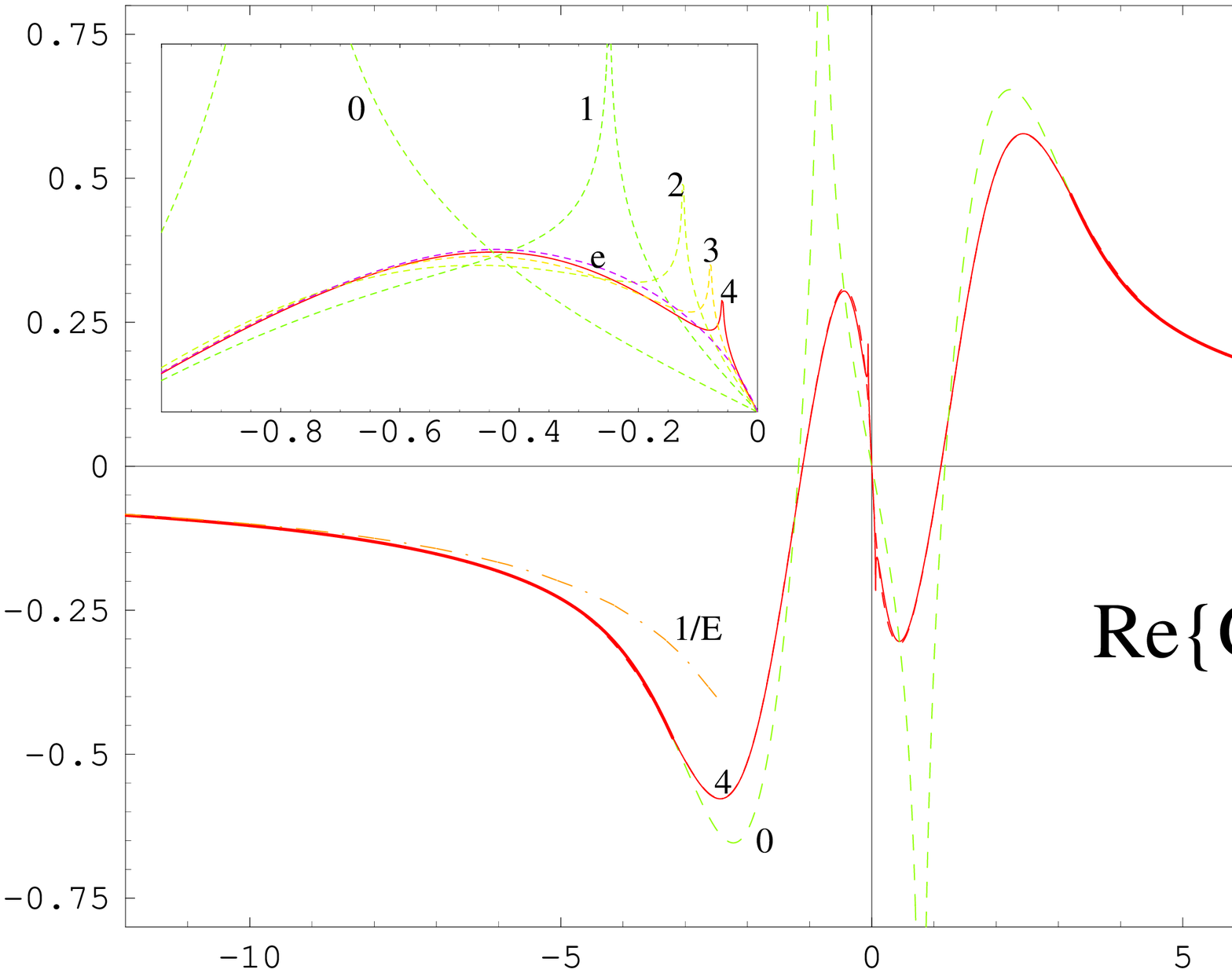}
}
\caption{The real part of the fermion Green function is shown for $0$- and $4RSB$
over a wide energy range. The zoomed region (upper left inset) shows all calculated
$RSB$-orders ($0,1,2,3,4$). The result from Eq.\ref{spectralrep} using the
$\infty RSB$-DOS model is contained (dashed curve carrying label e)
and visibly different from $4RSB$ only in the zoomed inset.
Gap edge singularities (spikes) with positions depending on $RSB$-order $k$
are observed; curves are labeled by the order $k$.
The logarithmic gap edge singularities are cut off in numerical calculations
by small imaginary parts $\delta=O(10^{-6})$ in $E+i\delta$ which ensure a finite
distance from the branch cut of $G$. The free $1/E$-behaviour is
shown for comparison.
}
\label{Fig:reg20}
\end{figure}
Figure \ref{Fig:reg20} (inset) shows the step-by-step approach of the exact
solution. The limit of spin glass interaction effects is shown by the crossover
into the free limit $1/E$; together with this one observes an energy range where
even the replica symmetric approximation is almost exact (roughly for $|E|>3$).
The inset shows the increase in length of the well-approximated energy
regimes with the growing order of $RSB$. The Figure also confirms the existence
of hierarchical excitation energy regimes, where a certain minimal order of $RSB$
guarantees an almost exact result. Moreover these regimes appear to scale with
corresponding characteristic temperature regimes.
We further observe that the logarithmic divergence of $Re(G)$ at the Fermi level
appears to be compatible for the scale chosen for the main Figure; in the zoomed
inset the slope at $E_F$ looks finite even in the interpolation model. This model
however includes the logarithmic divergence which emerges only in an
exponentially small (invisible) energy range near the Fermi level.\\
The analytical result in $0RSB$ for the retarded fermion Green function for $T=0$
and half-filling reads
\begin{equation}
G^R(E)=\frac{1}{\sqrt{8\pi}}\sum_{\lambda=\pm}\left[\lambda\hspace{.1cm}
exp(-\frac12 c_{\lambda,E}^2)\Gamma(0,-c_{\lambda,E}^2)-i\pi
exp(-c_{\lambda,E}^2)(1-erf(-i\hspace{.1cm}c_{\lambda,E}))\right]
\label{0rsb_G}
\end{equation}
where $c_{\lambda,E}=E+\lambda\bar{\chi}+i0_{+}$. The
$\Gamma$-function in this expression reproduces the logarithmic
singularities at the gap edge energies as seen in the numerical
calculation above. Eq.\ref{0rsb_G} crosses over into the free
$1/E$-behaviour for large energies and is almost exact beyond
$|E|>3$, ie for excitation energies more than four times beyond
the spin glass temperature (recalling $J=1$). From
Eq.\ref{spectralrep} it is clear that whenever $\rho(u)$ drops to
zero at a gap edge energy, the Cauchy principal value integral
must diverge logarithmically. Only in the $\infty RSB$ limit is
$\rho$ no more discontinuous; it is even in $u$ and no divergence
occurs in $Re(G(0))$. For finite Hubbard interaction the gap is
not closed, but the density of states decays at least with some
power at the Hubbard gap edge, which again prevents a singularity
in $Re(G)$. By means of the rapid convergence of $RSB$, the
numerical analysis supports that the slope of $G(E)$ at $E=0$
diverges. Only in case when the spectral density vanishes faster
than linearly at $E=0$ the slope would remain finite.\\

Figure 21 (fig21.gif) appended at e.o.f.
\begin{figure}
\caption{Part a) shows $Re(G(\epsilon))$, $\epsilon$ real, in 4RSB-approximation
and superimposed the result obtained from the interpolation model for the exact
density of states. Very good agreement is obtained. Both curves can hardly be
distinguished except in the tiny regime near the small 4RSB-gap edge at
$|\epsilon|\approx0.06$, which is absent in the exact solution.
Innocuous too, although present in the interpolating model solution in an
exponentially small region near the Fermi level, is the logarithmically divergent
slope of the real part of $G(\epsilon)$.\\
Part b) shows the effect of the Hubbard interaction for $U=1$ (in units of $J$)
in comparison with the $U=0$-result contained in part a).
For $U=1$ the slope of $Re(G)$ does no more diverge at the Fermi level.
The Hubbard interaction squeezes a hard gap of size $U$ into the spin glass
pseudogap. The logarithmic divergence at the Fermi level is
therefore removed for all finite $U$.}
\label{Fig:reg2}
\end{figure}
In Fig.\ref{Fig:reg2} part b) a solution for finite Hubbard interaction $U=1$ is
compared with the $U=0$ case; we justify the use of the $\infty RSB$-model by
the very good agreement with the calculated $4RSB$-curve (once the $RSB$-artifact
at the gap edge is omitted). The density of states together with Fig.\ref{Fig:reg2}
give a complete picture of the true (local) fermion propagator.
Starting from the numerical data for arbitrary values of the Hubbard coupling
(in units of the spin glass coupling $J$) we are ready to create,
on the basis of computer algebra, objects that can be dealt with
like standard Green functions and hence be used in diagram theories.

\begin{figure}
\centerline{
\epsfxsize=11.5cm
\epsfbox{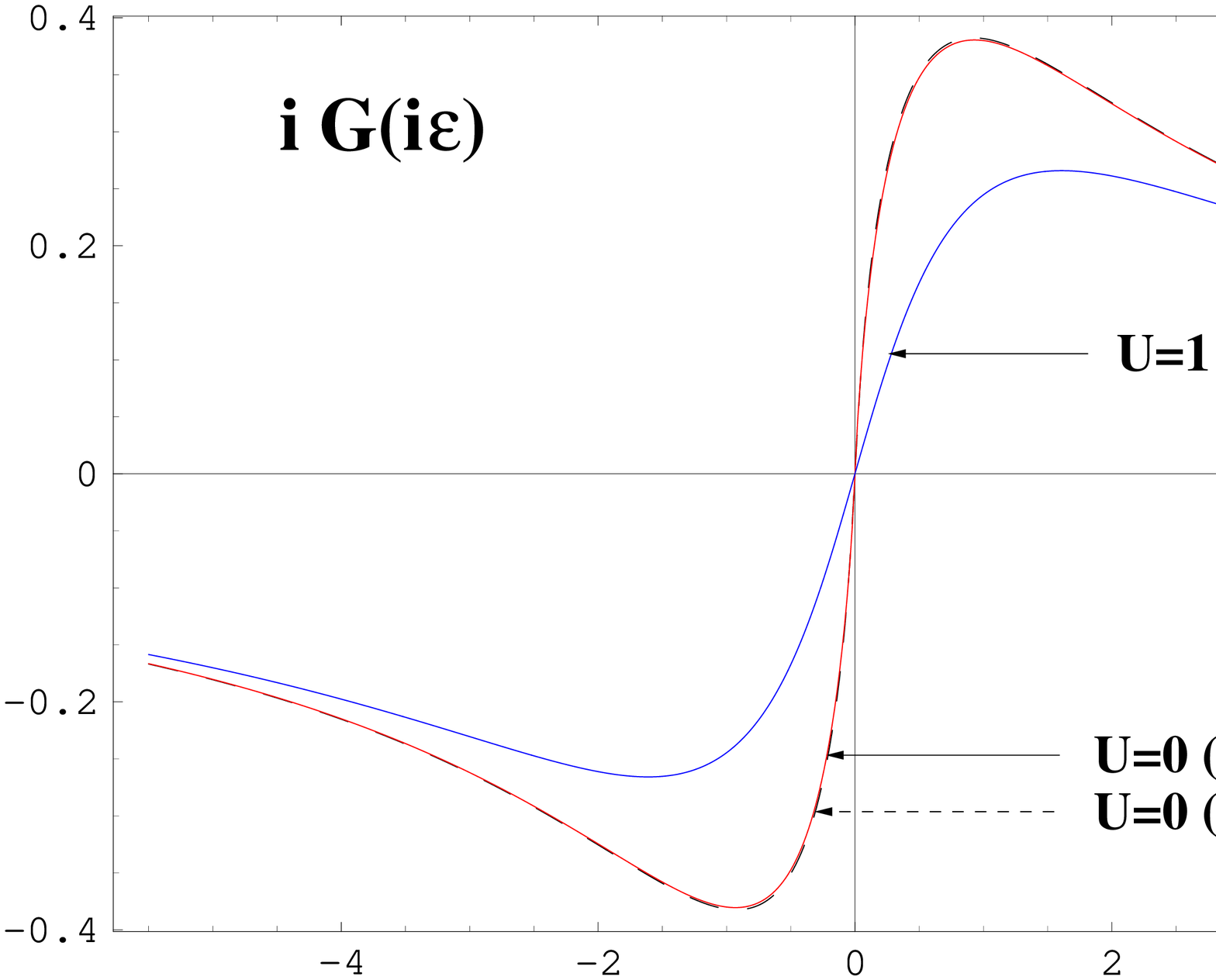}
}
\caption{The (imaginary) fermion propagator $G(i\epsilon)$ for half filling
is displayed as a function of the fermionic frequencies
$z_n=(2n+1)\pi k_B T/\hbar\rightarrow_{T\rightarrow0}\epsilon$
which become continuous in the zero temperature limit.
For $U=0$, the dashed curve represents the 4$RSB$-calculated one, which almost
perfectly coincides with the one calculated from the DOS-interpolation model for
full $RSB$.
The Hubbard interaction effect is shown for the chosen strength $U/J=1$.}
\label{matsubaraG}
\end{figure}

The numerical analysis is completed by Figure \ref{matsubaraG} for the purely
imaginary Green function $G(iz_n)$.
Figure \ref{matsubaraG} shows a spectacular convergence already in the available
low orders of $RSB$. Specific singular features, observed in $G^R(E)$
for real energies $E$, are absent on the imaginary axis (of course
they are hidden and almost innocuous in $G(iz_n)$)
The absence of singularities on the imaginary energy axis gives rise to the
speculation that an unexpected simple approximate form of the exact spin glass
propagator may be good enough for calculations on the imaginary axis
(leaving aside problems with the final analytic continuation to real energies).
\subsection{Scaling at the half-filling $T=0$-transition}
The half-filling transition is sharp at $T=0$ and becomes continuous as $\bar{\chi}$
vanishes in the $\infty RSB$-limit. It can be driven by the variation of
either $\epsilon=E-\mu=E-E_F$ which controls the pseudogap-shape, or by $U-E_F$
which breaks down the central band. Finite temperatures act like a symmetry breaking
perturbation which smears the transition.
Our detailed $RSB$-calculations suggest the following scaling behaviour
\begin{eqnarray}
& &\rho(\epsilon,T=0,\mu<U)\approx a_1|\epsilon|+
A_{\alpha}|\epsilon|^{\alpha}\hspace{1cm}{\rm\hspace{.2cm}
gap\hspace{.2cm} regime}\\
& &\frac{dG}{d\epsilon}
\sim ln|\epsilon|\hspace{.4cm}{\rm for}\hspace{.2cm}\epsilon\rightarrow0\\
& &\nu-1\sim(\mu-U)^2\sim
1-q^{aa}=1-q_1
\hspace{.3cm}{\rm
for\hspace{.2cm}}\mu>U
\end{eqnarray}
where Fig.\ref{nu-fit} suggests a subleading exponent $\alpha=2$.
Assuming that $\epsilon$ scales like $\mu-U$ near the transition one can propose
the scaling form for the density of states
\begin{equation}
\rho(\epsilon,\mu-U,T)=A\hspace{.1cm} |\epsilon|^x
f_{\rho}(\frac{\mu-U}{\epsilon},\frac{T}{\epsilon^y}),
\end{equation}
where, according to Eq.(40) our prediction is $x=1$ for the infinite-range model, and
$f_{\rho}$ is meant to be the scaling function of the density of states.\\
For any finite $k$-step RSB the transition is discontinuous; viewing the
influence of smaller and smaller order parameters $q_k$ as $k\rightarrow\infty$ as a
disorder fluctuation effect reminds of conclusions stating
that disorder fluctuations can render a transition continuous \cite{cardy,berker}.
The difference would be that we are concerned here with a one-dimensional effect
in replica space, while those authors referred to real space fluctuations.
%
%
\section{A scenario for the spin glass driven metal-insulator transition}
%
\label{sec:MIT}
\subsubsection{Non-half-filled system}
According to the preceding chapters, a central band always exists if the system
is not half-filled. Then, the Fermi level must lie inside the upper spin glass
pseudogap, which is separated from the Hubbard gap by the central band.
A metal insulator transition must take place as fermion hopping increases
beyond a critical value. Two possibilities arise: first the pseudogap
defined by
$\rho(E_F)=0$ survives until the metal insulator transition ($MIT$) takes place or,
more likely, the pseudogap is gradually filled but states remain localized until
the density of states becomes large enough. Then the $MIT$ would take place.
Its character must be different from standard Mott-Hubbard-Heisenberg metal
insulator transitions,
since the gap's existence rests exclusively on the frustrated magnetic interaction
and, in addition, the localization of states stems again only from the randomness
of this interaction. No further random potentials or random scattering is necessary
to localize states in a way similar to Anderson localization.
We explained already in the introduction how the random magnetic order might act in
this way.
In this context one should recall that broken time reversal allows for the
unitary type of Anderson localization.
In this respect it is an interesting detail that spin glass order is
a special breaker of time-reversal invariance $TRI$, since the random
magnetic moments break $TRI$ locally, but globally the magnetization is zero
without field and a global average breaking is not present.
\subsubsection{Half-filling}
In the half-filled system all gaps combine into one. At $T=0$ the spin glass gaps
appear to be attached on each side of the Hubbard gap. The Fermi level lies in
the center and one could expect the metal-insulator transition to be similar
to that of the Hubbard model. However a detailed analysis must scrutinize this,
in particular when the variance of the frustrated interaction is larger than
the Hubbard coupling $U$.
\begin{figure}
\centerline{
\epsfxsize9.cm
\epsfbox{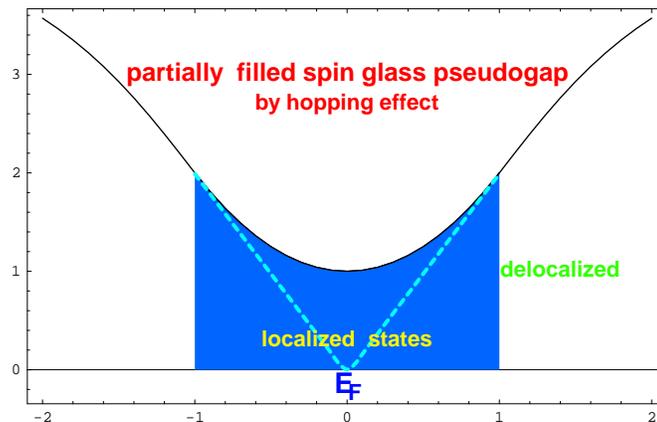}
}
\caption{Expected weak filling of the spin glass pseudogap by fermion hopping in non
half-filled systems; the shown form is assumed and not yet based on calculation.
Anderson-like localization, generated by the random magnetic interaction,
is indicated in the small density of states regime.
The dashed line sketches the possibility of critical depletion at the metal-insulator
transition as in the Coulomb case.}
\label{fig:MIT}
\end{figure}
\section{Fermionic SK-models with Hubbard-$U$ distributions}
%
\label{sec:randomU}
This paper has focused so far on effects of fixed Hubbard couplings and
the final section can neither give a comprehensive overview nor repeat with
the same depth $RSB$- and Green function-calculation for $U$-distributions.
We limit the detailed analysis of random-$U$ models to effects concerning
the freezing temperature, including modifications of tricritical behaviour
We discuss related physical problems which necessitate further studies in the future.
A lot of interesting possibilities emerge in models with random distributions of
different Hubbard-couplings or with randomly placed Hubbard centers.
Examples include long-standing physical problems like the ones mentioned in the
introduction and described in Refs.\onlinecite{mihill.shrgtn,hertz}, as well as
new problems which appear in artificially tailored systems, or in mathematical
toy models which help to probe the spin glass state by means of different kinds of
$U$-distributions. Apart from randomness, particularly interesting generalizations
are sublattice structures modeling A-B alloys, for example with alternating signs of
$U$, which allow staggered magnetic order in the sense of staggered deviations
of the spin glass order parameter from its mean value.
Such insulating states are nevertheless highly pregnant with competing or perhaps
coexisting magnetic superconductivity (or superconducting magnetism)
if only hopping processes are taken into account.
Ioffe and Larkin \cite{ioffe.larkin} described disordered superconductors with
smeared transition temperatures and randomly distributed couplings. They
considered percolation of clusters, in which superconducting order can resist the
destructive proximity effect of surrounding non-superconducting material.
Below we consider a few different model classes which, according to what was
achieved in the previous sections, shed some light on the interplay between
randomness in $U$ and the freezing temperature. Distributions of Hubbard couplings
of the site-uncorrelated Wilson-Ising potential
$$P(\{U_i\})={\cal{N}}^{-1}exp(-a_{_2} U_i^2 - a_{_4} U_i^4)$$ facilitates the
crossover from standard Gaussian distributions to a two level $\delta$-distribution
(random AB alloy). One may include site-site correlations by adding an additional
Wilson-like term $$exp(-\frac12 (\nabla_i U_{i})^2)\hspace{.5cm},$$
where either $\nabla$ denotes the discretized gradient on the lattice or one may
think of the continuum limit.
We will restrict the discussion to cases of a correlation-free distribution,
which will not invalidate the chosen form of the spin glass order parameter.
%
\subsection{White noise distributed Hubbard couplings $U_i$}
%
We study first the case of a local Gaussian disorder correlation (white noise)
described by
\begin{equation}
\langle (U_i-\langle U_i\rangle)(U_j-\langle U_j\rangle)\rangle=\delta U^2 \delta_{ij}
\end{equation}
The $U$-average of the replicated partition function considered by
Eq.(\ref{partition.function}) can be written as
\begin{equation}
\prod_i\int_{-\infty}^{\infty}\frac{dx_i}{\sqrt{2\pi}}e^{-x_i^2/2}
\int {\cal{D}}\Psi exp\left[\frac14\sum_i (U+\delta U\hspace{.1cm}x_i)\epsilon
\sum_{a,k}(\tilde{n}_{iak}^2-\tilde{\sigma}_{iak}^2)\right]{\cal{R}}\hspace{.2cm},
\end{equation}
where ${\cal{R}}$ stands for the $U$-independent part in the partition function
of Eq.\ref{avZ2}; $\tilde{n}$ and $\tilde{\sigma}$
are shorthand notations for charge and spin operators expressed in terms of
Grassmann fields, which are integrated over by means of
$\int{\cal{D}}\Psi$ (expression $\tilde{n}^2-\tilde{\sigma}^2$
coincides with the square bracket of Eq.5).
One can see that for Gaussian distributions what one needs to do is to replace $U$ by
$U+\delta U x_i$ and to perform the $x_i$-averages at each site.\\
The averaged freezing temperature can then be found from
\begin{equation}
T_f=\int_x^G
1/\left[1+exp(-\frac{1+U+\delta U x}{2T_f})
cosh(\frac{2\mu-U-\delta U x}{2T_f})\right]
\label{Eq:randomTf}
\end{equation}
By the help of $U$-averaged selfconsistent equations the following problems can
thus be answered immediately:\\
i) change in freezing temperature,\\
ii) can continuous transitions occur down to zero temperature,\\
iii) how strong is the depression of spin glass order and of $T_f$ by attractive
interaction (negative $U$) in comparison with the restoring repulsive interaction
(positive $U$); distributions can be defined which probe this
competition by the relative weight, the limit of infinitely broad
Gaussian $U$-distributions being a special case, which should render
the mean coupling $\langle U\rangle$ irrelevant, and\\
v) can random alloy models with special permitted $\langle U\rangle$-values
be expected to exhibit significant behaviour?

\begin{figure}
\centerline{
\epsfxsize=8cm
\epsfbox{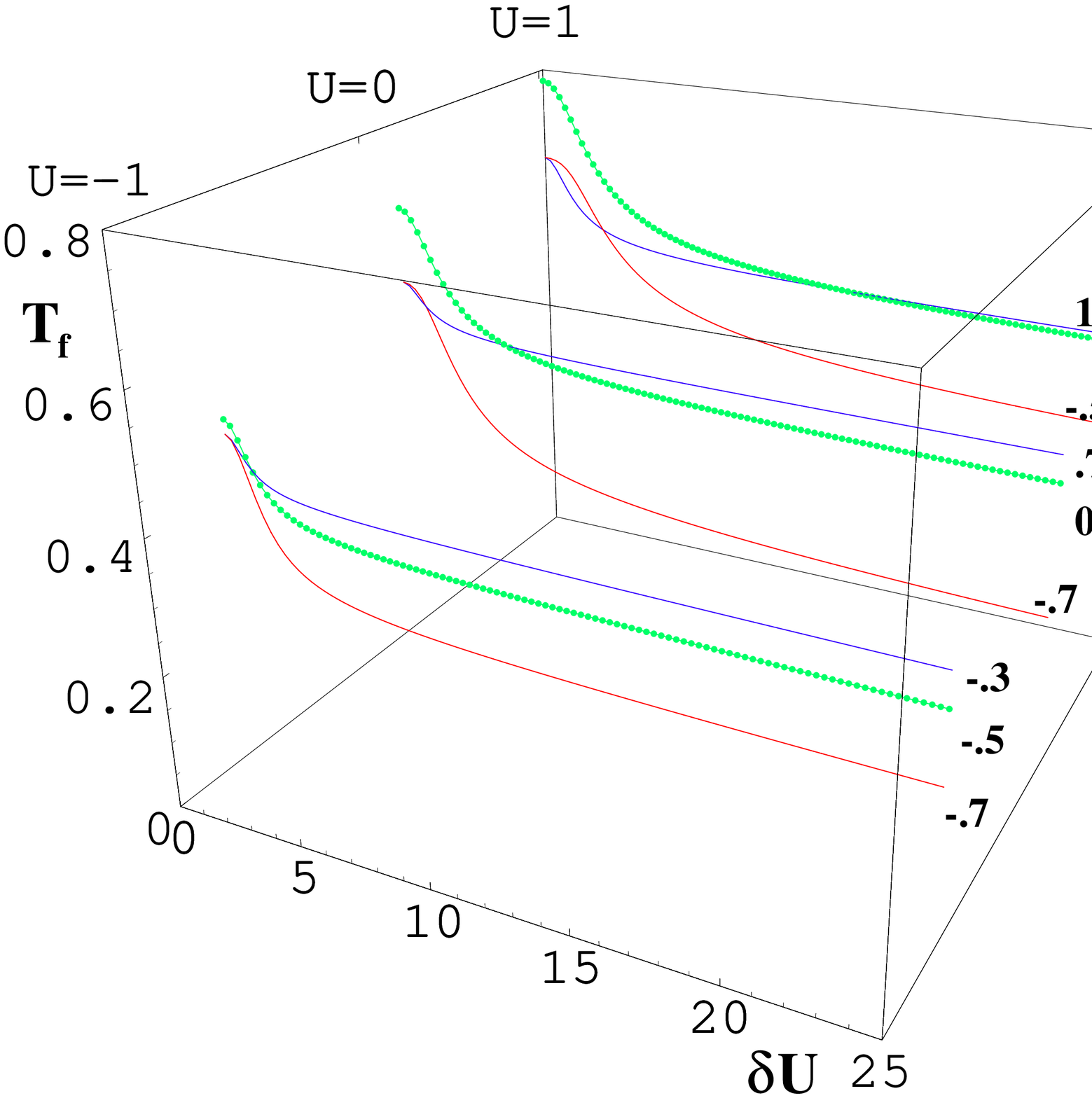}
}
\caption{
The variation of the freezing temperature $T_f$ under increased variance
$\delta U$ of the Gaussian $U$-distribution is shown for three chosen values
of the Hubbard coupling, $U=-1$ (attractive), $U=0$, and $U=1$ (repulsive).
For each $U$, $T_f(\delta U)$ is shown for three different chemical potentials
($\mu$-value given at the right end of $T_f$-curve), which are symmetrically
placed around the half-filling condition $\mu=U/2$ for $\delta U=0$.
}
\label{randomU}
\end{figure}

\begin{figure}
\centerline{
\epsfxsize=8cm
\epsfbox{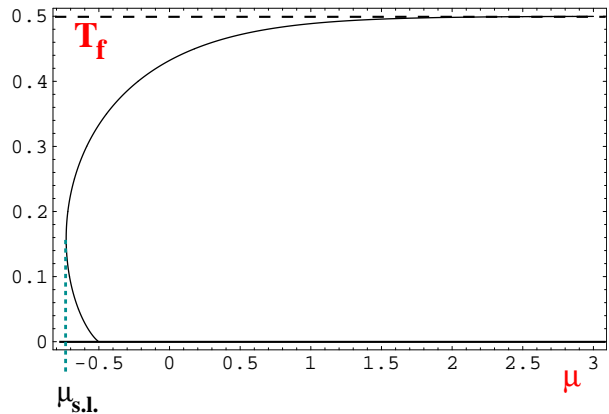}
}
\caption{Solution of $T_f$ in the limit of infinitely broad Gaussian distribution of
$U$: one finds $T_f=\frac12$ in the limit of large $\mu\rightarrow\infty$
while, in the small $\mu$-regime, a stability limit shows up at $\mu_{s.l.}=-0.7315$
and implies a discontinuous drop to $T_f=0$.
}
\label{Tflimit}
\end{figure}

According to Eq.\ref{Eq:randomTf}, $\delta U$ cannot be absorbed in a $\mu$-shift.
The asymmetric $U$-dependence (attractive $U$ suppress spin glass order
while repulsive interactions support it) allows a crossing of the critical curves
$T_f(\delta U)$.
As one can further observe in Fig.\ref{randomU}, the critical temperature
$T_f(\delta U)$ averages over the competing effects of positive and negative $U$ and
varies little for broad distributions.
The selfconsistency equation for $T_f$ is simplified for
$\delta U\rightarrow\infty$ and reads
\begin{equation}
T_{f}=1/\left[2+e^{(-\mu-1/2)/T_{f}}\right].
\end{equation}
which is $U$-independent because of the infinitely broad distribution. The
freezing temperature approaches $T_{f}(\mu)=\frac12$ for
$\mu\rightarrow\infty$ but drops discontinuously to zero near the stability limit at
$\mu_{s.l.}=-0.7315$ as shown in Figure \ref{Tflimit}.
The limits $\delta U\rightarrow\infty$ and $\mu\rightarrow\infty$
do not commute. For fixed $U$-variance and arbitrary $U$ one finds that the freezing
temperature decreases exponentially for $\mu\rightarrow\infty$.

\begin{figure}
\centerline{ \epsfxsize=9.5cm
\epsfbox{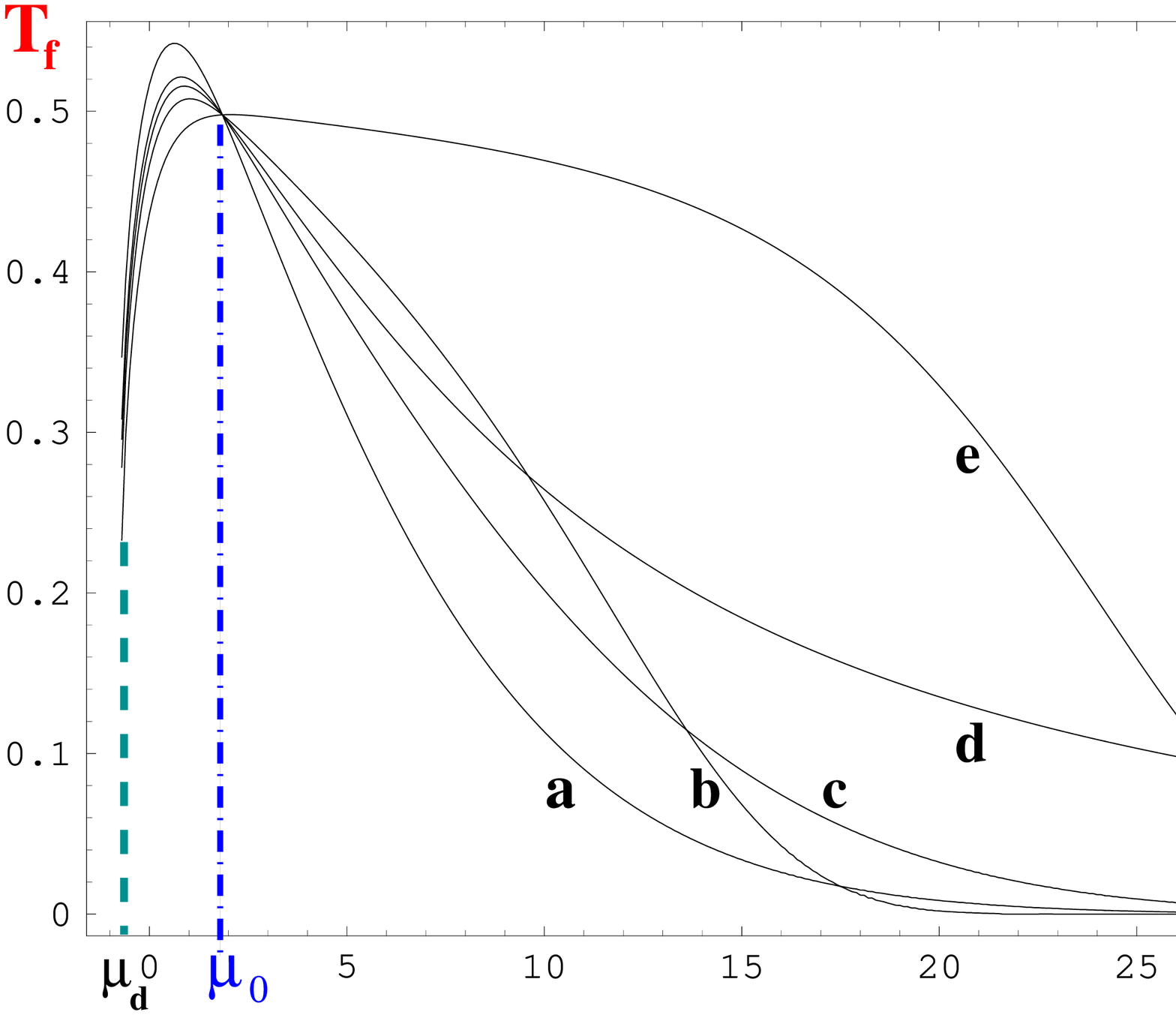}
}
\caption{Exponential decay of $T_f$ for the normalized $\delta U$-distributions
a) $\sim exp(-x^2/2)/(x^2+1)$, b) $\sim 1/(x^2+1)$, c) $\sim exp(-x^2/2)$,
d)$\sim exp(x^2-x^4/2)$, and e) $\sim exp(x^2-x^4/10)$ with fixed $U(=1)$.
}
\label{Udistribs}
\end{figure}

For several differently chosen distributions the averaged $T_f$-equation results
in Figure \ref{Udistribs} showing an exponential decay in the large $\mu$-limit
of the critical temperature (for fixed $U$). A universal temperature
$T_{f0}\equiv T_f(\mu_0)$ is observed, where $T_{f0}$ is independent
of distributions which are symmetric with respect to $U$.
The value $\mu_0$ of this invariant point depends linearly on $U$ and obeys
$\mu_0(U)=0.9939(U+0.8626)$.
One may check that the critical temperature $T_f$ given by Eq. \ref{Eq:Tf} and
Figure \ref{Tf} at $\mu_0$ agrees too.
For negative $\mu$ the freezing temperature vanishes discontinuously.
A line of tricritical points exists.
The exponential tails at large positive $\mu$ originate in the rare large $U$-regions.
None of the given examples allows a continuous phase transition at $T=0$, since $T_f$
remains nonzero for all finite $\mu$ (allowing only a discontinuous drop to $T_f=0$).
%
\subsection{Alloy models with two different Hubbard couplings}
%
Random alloys with only two different Hubbard couplings $U_1$, $U_2$, which are
assumed to be realized with probability $w_1$ and $w_2=1-w_1$ respectively on
each site obey
\begin{equation}
P(U_i)=w_1\delta(U_i-U_1)+(1-w_1)\delta(U_i-U_2)
\end{equation}
In this model the spatial homogeneity of the spin glass order parameter
$\langle Q\rangle$ remains unaffected. We observe that the
$T_f$-solution for this model also passes through the universal
point of Figure \ref{Udistribs} provided $w_1=\frac12$.
\\
A highly interesting $AB$-alloy model with staggered Hubbard-interaction,
defined with a $U$-distribution
\begin{equation}
P(\{U_i\})=\delta(U_i-U_A)\delta_{i,i_A}+\delta(U_i-U_B)\delta_{i,i_B},
\end{equation}
can break the spatial homogeneity of $\langle Q\rangle$.
In case that $U_A$ on sublattice $A$ is strongly negative, this sublattice
will become nonmagnetic, while the range-free magnetic interaction maintains
glassy order on sublattice $B$ provided $U_B$ is repulsive for example.
When nearest neighbor hopping of sufficient strength is
introduced, an extreme proximity effect between superconductivity
and magnetism will result: local pairs can only become delocalized
when single fermions tunnel through the magnetic sites. A strong
filling dependence must be expected. In addition the strong interference
of an eventually superconducting sublattice system with glassy magnetic order
is a challenge for research on microscopic superconducting glass
phases.
%
\section{Antiferromagnetic- and ferrimagnetic glassy phases}
\label{section:AFM}
The simplest model with range-free magnetic interaction, which nevertheless supports
antiferromagnetic order, is the Korenblit-Shender model \cite{korenblitshender}.
The magnetic interaction is supposed to act only between $A$ and $B$ atoms,
which one may imagine to sit on neighbouring sites, with a mean antiferromagnetic
form together with random fluctuations.

In previous work \cite{osk} we solved this model in one-step $RSB$ and generalized
it in several directions. Spin glass order parameters $Q_A$, $Q_B$ within each
sub-lattice differ in the glassy antiferromagnetic phase transition, and a third
field is necessary to describe the $A-B$ coupling. In a field we found a level
crossing of elements of $Q_A$ and $Q_B$, which led us to define a new $RSB$-scheme.
This scheme is allowed to be $A-B$ asymmetric which, in the strongly $A-B$-asymmetric
ferrimagnetic phase, turns out to be better than the standard solution. Details will
be published elsewhere.\\
The Hubbard interaction can be built into the antiferromagnetic model in the same way
as described before. Hubbard interactions $U_A$, $U_B$ of different strength can
enhance the $A$-$B$ asymmetry. This is evident from the fact that sufficiently
negative $U$, say on $A$-sites, render these sites nonmagnetic, while sites with
positive $U$, say on $B$-sites, maintain glassy order. This case has very strong
asymmetry and the new $RSB$-scheme of Ref.\onlinecite{osk} must be considered as
a candidate to describe such phases.\\
Such problems are left for future research. When hopping is also allowed for,
superconductivity on the $A$-sublattice in competition with glassy magnetism
on $B$-sites revive the problem of microscopic superconducting glasses as well.
\section{Summary of results and outlook}
\label{section:summary}
We presented the analytical solution for the generating functional of a model with
competing Hubbard- and frustrated range-free Ising spin interaction
together with detailed numerical evaluations of the resulting self-consistency
equations in order to describe relations between magnetism and electronic properties.
The model was interpreted as the localized (but non-local) limit of a Hubbard model
with additional frustrated spin interaction.
The discrete time slicing method was, to our knowledge, used for the first time
to solve a spin glass problem analytically (all integrations were performed before
the continuum time limit was taken).
Taking advantage of the rapid convergence of $k$-step replica symmetry breaking
and using exact relations we drew conclusions on the $\infty RSB$ solutions
of important quantities such as the fermion propagator $G(E)$.
The $2-4$-step approximations were found to be close enough to the exact solution
(by means of comparison with exact relations) and allowed us to identify and to
eliminate artifacts of finite-step RSB in the extreme low temperature regime.
By comparison of $0$-,$1$-,$2$-,$3$, and $4$-step results, we identified hierarchical
energy regimes, $\overline\epsilon^{(k)}<|\epsilon|<\underline{\epsilon}^{(k)}$,
where corrections beyond a certain $k$-step-approximation are negligibly small.
For decreasing excitation energies above or below the Fermi level an
increasing number of steps is required at $T=0$ for good accuracy.

We found that all $T=0$ phase transitions of the model (including its generalization
to random distributions of the Hubbard coupling) are discontinuous with the
exception of half-filling transitions. Those become continuous in the limit of
infinite-step RSB (i.e. their discontinuities decay with increasing number of
$RSB$-steps and vanish in the limit of $\infty RSB$).

The density of states around the spin glass pseudogap at $E_F$ was found to
obey the linear dependence $\rho(E)=c_1 |E-E_F|$ with $c_1\approx0.3$ in a
wide gap regime and $c_1\approx0.13$ asymptotically close to the Fermi level.
This was concluded from a fast convergence of $\rho'(E)$ under increasing number
of $RSB$-steps. The linear behaviour of $\rho$ reminded us of the
Efros-Shklovskii Coulomb-gap in two dimensions.
These gaps have totally different origin: the Efros-Shklovskii gap
originates in the bare Coulomb interaction (and naturally involves a dependence
on space dimension) while the present one is caused by the spin glass order.
In finite-range models the latter is expected to depend on space dimension
predominantly through the destruction of glassy order at its lower
critical dimensions for finite temperature and $T=0$-transitions
respectively.

We believe that a renormalization group analysis acting in replica space
(not in real or momentum space) will be helpful to describe the asymptotic behaviour
linked to the hierarchical energy scales, which were discussed in our
paper for the range-free model. Eventually a new type of RG must be
created for this purpose.
In accordance with general considerations \cite{shankar},
Johnson and Khmelnitskii \cite{khmelnitskii} showed
for a disordered system with long-range (unscreened) Coulomb interaction
how important and useful a RG-technique can be for the understanding of
pseudogaps.


From our numerical analysis we predicted a set of other scaling laws.
The linear decay of the density of states led us to claim a divergent slope of the
Green function's real part at $E_F$ and hence a divergence of a corresponding vertex,
which results from charge conservation Ward identity \cite{ro.epl}.
These divergences are logarithmic and become visible only exponentially close
to the Fermi level.
This reminds of the extremely slow approach to equilibrium, which is usually
discussed for spin dynamics. In the present case, in the absence of spin dynamics
it is the quantum fermion dynamics, generated by the noncommuting fermion- and
Hamilton operator, which experiences the hierarchical $RSB$-structure of spin glass
order parameters.
For the fermion concentration (integrated density of states at $T=0$) and for
the related $T\rightarrow0$ limit of the integrated Parisi function,
$q_1=\lim_{T\rightarrow0}\int_0^1 dx\hspace{.1cm} q(x,T)$,
we predicted $\nu-1\sim 1-q_1\sim(\mu-U)^2, \mu>U$, near the half-filling transition.
The transition generates an important crossover of the Fermi level from the center
of a spin glass pseudogap into the Hubbard gap and hence into a deeper insulating
state. We stressed its role for delocalization in itinerant models and described
related scenarios for metal insulator transitions.

The present approach was intended to progress towards an analytical solution
at $T=0$ and the construction of a low temperature expansion.
A generalization to fermionic spin glasses of a recent numerical study by
Crisanti et al\cite{crisanti}, derived from an optimization algorithm and
analogy with $T=0$ standard spin glasses, could be a new alternative route
way as well as the application of a renormalization group in replica space.

The models we studied discarded the long-range part of the Coulomb interaction.
This interaction, which is known to produce the Efros-Shklovskii Coulomb pseudogap,
should be taken into account in a more complicated model.
The eventual combination with the spin glass pseudogap should be analyzed.

We finally note that the smooth form obtained for the fermion propagator
in section \ref{fermion-propagator}, and its analytical approximation,
can be a starting point of a perturbation theory)
or at least control limit in studies of more complicated cases such as
itinerant models with fermion hopping, mixed valence (Anderson-)models, and s-d coupling models
between spins of mobile carriers and spin glass ordered localized spins.
%
\section{Acknowledgements}
This work was supported by the EPSRC under grant GR/R25583/01 and GR/M04426,
by the DFG under Op28/5-2 and SFB410,
and by the ESF-programme SPHINX. We wish to thank David Huse for helpful remarks.
One of us (R.O.) is grateful for hospitality extended to him at the department
of physics at Oxford university.


\begin{thebibliography}{99}
\bibitem{Diep.book}{\it Magnetic systems with competing interactions
(Frustrated Spin Systems)}, edited by H. Diep (World Scientific Publishing Co.,
1994)
\bibitem{fisherhertz} K.H. Fischer, J. Hertz, {\it Spin Glasses}
(Cambridge University Press, Cambridge, UK) 1991
\bibitem{mihill.shrgtn} D. Sherrington, K. Mihill, J.de Physique
{\bf 35} C-4, 199 (1974)
\bibitem{parisi} G. Parisi, J. Phys. {\bf A13}, 1101 (1980)
\bibitem{parisibook} G. Parisi, {\it Field Theory, Disorder and Simulations}
(World Scientific Publishing Co., Singapore) 1992
\bibitem{finkelstein} A.M. Finkelstein, Sov.Phys.JETP {\bf 57}, 97 (1983)
\bibitem{kotliar.et.al} A. Georges, G. Kotliar, W. Krauth, M.J. Rozenberg,
Rev.Mod.Phys.{\bf 68}, 13 (1996)
\bibitem{balachandran.review} A.P. Balachandran, E. Ercolessi, G. Morandi,
A.M. Srivastava, Int.J.Mod.Phys. {\bf B4}, 2057 (1990)
\bibitem{vollhardt} D. Vollhardt, in {\it Correlated Electron
Systems}, edited by V.J. Emery (World Scientific, Singapore, 1992)
\bibitem{metzner} W. Metzner, Phys. Rev. {\bf B43}, 8549 (1991)
\bibitem{laloux} A. Georges and M. Laloux, cond-mat/9610076
\bibitem{SK} D. Sherrington, S. Kirkpatrick, Phys.Rev.Lett.{\bf 35} 1972 (1975)
\bibitem{binderyoung} K. Binder, A.P. Young, Rev. Mod. Phys.{\bf 58}, 801 (1986)
\bibitem{ghatak} S.K. Ghatak, D. Sherrington, J.Phys.{\bf C 10}, 3149 (1977)
\bibitem{rohf} H. Feldmann, R. Oppermann, Phys.Rev.{\bf B62} 9030 (2000)
\bibitem{hertz} J. Hertz, Phys. Rev. {\bf 19}, 4796 (1979)
\bibitem{huse} O. Motrunich, S.-C. Mau, D.A. Huse, D.S. Fisher,
Phys.Rev.{\bf B61}, 1160 (2000)
\bibitem{osk} R. Oppermann, D. Sherrington, M. Kiselev, cond-mat/0106066
\bibitem{crisanti} A. Crisanti, L. Leuzzi, G. Parisi,
J.Phys.A:Math.Gen. {\bf 35}, 481 (2002)
\bibitem{remi} R. Monasson, R. Zecchina, Phys.Rev.{\bf E56}, 1357
(1997)
\bibitem{awschalom} I.P. Smorchkova, N. Samarth, J.M. Kikkawa, D.D. Awschalom,
Phys.Rev.Lett.{\bf 78}, 3571 (1997)
\bibitem{awschalom2} I.P. Smorchkova, N. Samarth, J.M.
Kikkawa, D.D. Awschalom, Phys.Rev.{\bf B58}, R4238 (1998)
\bibitem{negeleorland} J.W. Negele and H. Orland, {\it Quantum
many particle systems}, (Addison-Wesley, N.Y.) 1988
\bibitem{subir} A. Georges, O. Parcollet, S. Sachdev,
Phys.Rev.{\bf B63}, 134406 (2001)
\bibitem{cardy} J.L. Cardy, A.J. Mc Kane, Nucl.Phys.{\bf
B257[FS14]}, 383 (1985)
\bibitem{berker} A. Falicov, A.N. Berker, Phys.Rev.Lett.{\bf 76}, 4380 (1996)
\bibitem{aharony} F.C. Chou, N.R. Belk, M.A. Kastner, R.J. Birgeneau, A. Aharony,
Phys.Rev.Lett.{\bf 75}, 2204 (1995)
\bibitem{efros} A.L. \'Efros, B.I. Shklovskii, J. Phys. C{\bf 8}, {\bf L49} (1975)
\bibitem{khmelnitskii} S.R. Johnson, D.E. Khmelnitskii,
J.Phys.:Condens. Matter {\bf 8}, 3363 (1996)
\bibitem{dotsenko} V. Dotsenko, A.B. Harris, D. Sherrington, R.B.
Stinchcombe, J.Phys.A:Math.Gen. {\bf 28}, 3093 (1995)
\bibitem{shankar} R. Shankar, Rev.Mod.Phys. {\bf 66}, 192 (1994)
\bibitem{ro.epl} R. Oppermann, B. Rosenow, Europhys.Lett.{\bf 41}, 525 (1998)
\bibitem{ioffe.larkin} L.B. Ioffe, A.I. Larkin, Sov.Phys. JETP{\bf
54}, 378 (1981)
\bibitem{fedotov} V.N. Popov, S.A. Fedotov, Zh. Eksp. Teor. Fiz.
{\bf 94}, 183 (1988)[Sov.Phys.JETP {\bf 67}, 535 (1988)]
\bibitem{korenblitshender} I.Ya. Korenblit, E.F. Shender,
Sov.Phys JETP{\bf 62} 1030 (1985)

\end{thebibliography}
\end{document}